\documentclass[12pt]{article}
\usepackage{amssymb}
\usepackage{graphicx}
\usepackage{epsfig}
\usepackage{amsfonts}
\usepackage{axodraw}

\newcommand{\br}{\hskip .25cm/\hskip -.25cm}

\newcommand{\ol}{\overline}

\newcommand{\lapproxeq}{\lower .7ex\hbox{$\;\stackrel{\textstyle
<}{\sim}\;$}}
\newcommand{\gapproxeq}{\lower .7ex\hbox{$\;\stackrel{\textstyle
>}{\sim}\;$}}
\newcommand{\stackdown}[2]{\lower 1.4ex\hbox{$\;\stackrel{\textstyle{#1}}
{\scriptstyle{#2}}\;$}}
\newcommand{\be}{\begin{equation}}
\newcommand{\ee}{\end{equation}}
\newcommand{\beq}{\begin{equation}}
\newcommand{\eeq}{\end{equation}}
\newcommand{\bea}{\begin{eqnarray}}
\newcommand{\eea}{\end{eqnarray}}

\newcommand{\D}{\displaystyle}
\newcommand{\elle}{\ell\hspace{-0.16cm}/}

\newcommand{\pslush}{p\hspace{-0.16cm}/}

\makeatletter

\def\slash{\@ifnextchar[{\fmsl@sh}{\fmsl@sh[0mu]}}
\def\fmsl@sh[#1]#2{%
  \mathchoice
    {\@fmsl@sh\displaystyle{#1}{#2}}%
    {\@fmsl@sh\textstyle{#1}{#2}}%
    {\@fmsl@sh\scriptstyle{#1}{#2}}%
    {\@fmsl@sh\scriptscriptstyle{#1}{#2}}}
\def\@fmsl@sh#1#2#3{\m@th\ooalign{$\hfil#1\mkern#2/\hfil$\crcr$#1#3$}}
\makeatother
\def\beq{\begin{equation}}
\def\eeq{\end{equation}}

\catcode`\@=11 % This allows us to modify PLAIN macros.

\def\lsim{\mathrel{\mathpalette\@versim<}}
\def\gsim{\mathrel{\mathpalette\@versim>}}
\def\@versim#1#2{\vcenter{\offinterlineskip
    \ialign{$\m@th#1\hfil##\hfil$\crcr#2\crcr\sim\crcr } }}

\def\t1{{\tilde 1}}

\def\slash#1{#1\hskip-6pt/\hskip6pt}

\def\to{\rightarrow}

\begin{document}

\begin{titlepage}

\begin{flushleft}
CERN-PH-TH/2010-263\\
KCL-PH-TH/2010-32
\end{flushleft}
\vspace{0.2cm}
\begin{centering}

{\Large  \textbf{ Quantum-Gravity Induced Lorentz Violation and Dynamical Mass Generation }}
\vspace{0.6cm}

{\bf Nick~E.~Mavromatos$^{a, b}$}

\vspace{0.4cm}

$^a$ {CERN, Theory Division, CH-1211 Geneva 23, Switzerland.}

$^b$ {On leave from: King's College London, University of London, Department of Physics,
Strand WC2R 2LS, London, U.K.}

\vspace{0.6cm}

{\bf Abstract}

\end{centering}

\vspace{0.4cm}

In Ref. \cite{alexandre} a minimal extension of (3+1)-dimensional Quantum Electrodynamics
has been proposed, which includes Lorentz-Violation (LV) in the form of higher-(spatial)-derivative isotropic terms in the gauge sector, suppressed by a mass scale $M$. The model can lead to dynamical mass generation for charged fermions. In this article I elaborate further on this idea and I attempt to connect it to specific quantum-gravity models, inspired from string/brane theory. Specifically, in the first part of the article, I comment briefly on the gauge dependence of the dynamical mass generation in the approximations of \cite{alexandre}, and I propose a possible avenue for obtaining the true gauge-parameter-independent value of the mass by means of Pinch Technique argumentations. In the second part of the work I embed the LV QED model into multibrane world scenarios with a view to provide a geometrical way of enhancing the dynamical mass to phenomenologically realistic values by means of bulk warp metric factors, in an (inverse) Randall-Sundrum hierarchy. Finally in the third part of this note, I demonstrate that such Lorentz Violating QED models may represent parts of a low-energy effective action (of Finsler-Born-Infeld type) of open strings propagating in quantum D0-particle stochastic space-time foam backgrounds, which are viewed as consistent quantum gravity configurations. To capture correctly the quantum fluctuating nature of the foam background I replace the
D0-recoil-velocity parts of this action by appropriate gradient operators in three-space, keeping the photon field part intact. This is consistent with the summation over world-sheet genera in the first-quantized string approach.
I identify a class of quantum orderings which leads to the LV QED action of Ref.~\cite{alexandre}. In this way I argue, following the logic in that work, that the D-foam can lead to dynamically generated masses for charged-matter (fermionic) excitations interacting with it.

\end{titlepage}

\setcounter{equation}{0}

\section{Introduction}

Lorentz Violation (LV) in the Standard model is a relatively old subject, originating from works by Kostelecky and collaborators~\cite{kostel}. There is an extensive literature over the past twenty years on tests of LV based on a (phenomenological) effective Lagrangian approach, the so-called Standard Model Extension (SME). However, there is no way in this approach of knowing the precise magnitude of the various LV parameters appearing in the SME, and thus we cannot have a feeling on the expected order of magnitude of the violations, so as to guide (or discourage!) experimental searches in a sensible and constructive way.

It is therefore desirable to have an ultraviolet completion of such phenomenological theories, which would guide us in our quest for an understanding of the quantum structure of space time at miscroscopic scales. In this latter respect, Horava recently re-visited the issue of Lorentz violation by suggesting an anisotropic time-space scaling for gravity, which in the ultraviolet regime rendered the quantum theory of this gravitation renormalizable~\cite{horava} (Horava-Lifshitz theories). This approach triggered an enormous interest
among the relevant communities. Although in its original formulation the model required a cosmological constant with the wrong sign to be compatible with observations, nevertheless, phenomenologically realistic improved versions~\cite{sotiriou} and appropriate extensions of the Standard Model~\cite{posp} have recently been constructed. Moreover, LV quantum field theories with anisotropic scalings (therefore of Lifshitz type, but not related necessarily to gravity) existed long before the Horava-Lifshitz gravity, and in fact their anisotropic scaling has been used as a regulator of the ultraviolet divergences of quantum field theory~\cite{visser}.
Dynamical mass generation in such theories has also been discussed~\cite{dynmas}.

Whether LV is ultimately related to quantum gravity is not known at present, since all such theories are at the level of effective (low-energy) field theories and their microscopic understanding is lacking.
However, Quantum Gravity may not admit a local effective lagrangian description, as we have argued some time ago, based on a toy model of space-time foam, originating from string theory~\cite{dfoam,westmuckett,emnnewuncert,li}. In fact, in these works, the space-time foamy structures are provided by stringy defects, interacting with matter string excitations on brane Universes.
The defects break Poincar\'e invariance, and their recoil during the interaction with open strings, also breaks local Lorentz invariance. The D-particles, being string theory solitons, quantum fluctuate
and their fluctuations give the background space time a foamy nature (``\emph{D-foam}'').
There is a non trivial momentum transfer during the interactions of matter strings with D-foam defects, which necessitates the emission of non-local intermediate string states. The latter do not admit a local effective action description. In this kind of models, the so-called ultraviolet completion is provided by String/Brane theory itself. We have argued~\cite{emnnewuncert,review}, however, that, despite the lack of an efficient local effective action description, there is a rich astro-particle phenomenology associated with either the induced vacuum refraction, characterizing the D-foam, or with global aspects of it, associated with modifications of the energy budget of the Dark sector of the Universe~\cite{westmuckett,emnnewuncert,vergou}.

Nevertheless, as we shall discuss below, under some approximations, it is after all possible to write down a LV local effective action (expanded in powers of derivatives), which describes part of the effects of
an isotropic, stochastically fluctuating, D-foam on radiation and, indirectly, on charged fermionic matter coupled to it, in particular dynamical mass generation. However, there are some other aspects of the foam, such as the induced time delays~\cite{emnnewuncert}, during the interaction of a photon with a D-particle, and the associated vacuum refractive index, that such an action cannot fully capture. The latter effects, being associated with stringy time-space uncertainties, require the full string theory machinery to be quantified.

The quantum fluctuations of the D-particle defects in target space lead to a novel \emph{correspondence principle } by means of which a B-antisymmetric tensor background field in phase space, representing the recoil velocity of the defect during its interaction with matter, is mapped~\cite{szabo} into a spatial derivative operator along the direction of the recoil. In this way, the resulting Finsler-Born-Infeld (FBI) Lagrangian, which describes the (low-energy) dynamics of open strings on a brane world in interaction with the D-particles, may be transformed into an effective Lagrangian with Lorentz-violating higher spatial derivative terms. The latter are of the form studied in a minimal LV extension of QED considered recently in \cite{alexandre}. That model has been argued to provide  a novel way of generating electron masses dynamically. The model is not of Lifshitz type, in the sense that there is no anisotropic scaling between time and space coordinates. The presence of LV is manifested through higher-spatial derivative terms, respecting rotational invariance in three space, which are suppressed by an effective mass scale. The presence of this scale and the LV terms \emph{catalyze} dynamical mass generation for charged fermions, for arbitrarily weak gauge fields.

In view of our link of such a Lagrangian with the D-foam, and ultimately with more general quantum gravitational structures, this constitutes an explicit realization of the effects of a foam medium on `slowing' down some particles via mass generation. In our case, the suppression mass scale of the LV terms is expressed in terms of the string mass scale and fundamental parameters of the foam, such as the variance of its quantum fluctuations. There are issues associated with quantum ordering ambiguities in our construction, which I attempt to resolve by appealing to the important r\^ole of the dynamical mass generation in curing infrared instabilities in the model. In this way, I select as the \emph{physical} class of \emph{quantum orderings} the one that maps the FBI effective action into \emph{the precise form} of the LV QED action of \cite{alexandre} at low energies. Any further ordering ambiguities within this class of models is absorbed into the quantum fluctuation parameters of the D-foam, which at this stage are viewed as phenomenological.

The structure of the article is as follows: in the next section \ref{sec:2}, I review the basic features of the model of \cite{alexandre}. In section \ref{sec:3}, I comment briefly on the gauge dependence of the dynamically generated mass for fermions in this model, and propose a way to extract the gauge-fixing-parameter independent part of it, which we argue to be provided by its value in the Feynman gauge. The argumentation is based on the so-called pinch technique~\cite{Cornwall:1982zr,binosi2009}, which applies straightforwardly to our LV case, as a result of the very minimal form of the LV terms in the model of \cite{alexandre}. In section \ref{sec:4} I discuss geometric ways of enhancing the dynamically generated mass through an inverse Randall-Sundrum-like hierarchy, which is achieved when the model of \cite{alexandre} is embedded appropriately into a multibrane world scenario, with warped bulk geometries. As a rather intriguing aspect, in section \ref{sec:dfoam}, I discuss a possible microscopic origin of the LV terms in the model of \cite{alexandre}. I show that such terms may constitute parts of an effective action describing the interaction of photons with quantum fluctuating (in target-space) D-particles in the stochastic stringy space-time foam model of \cite{westmuckett,emnnewuncert}. In this case, the LV scale $M$ in the action of \cite{alexandre} is identified with a specific function of the foam parameters and the string scale and coupling. Conclusions and Outlook, in particular possible extensions of the model to incorporate flavoured neutrinos, are presented in section \ref{sec:5}. Details on the application of the pinch technique to our one-loop approximate Schwinger-Dyson (SD) equations, relevant for dynamical mass generation in the analysis of \cite{alexandre}, and some remarks on its current status in gauge theories in general, are given in an Appendix to the work.

\section{Review of a minimal Lorentz-Violating (LV) QED Model \label{sec:2}}

In \cite{alexandre} dynamical mass generation for fermions has been studied in the context of a (3+1)-dimensional QED field theory with higher-order spatial derivatives in the photon sector, that violated four-dimensional Lorentz symmetry but preserved spatial rotations, and a standard form for the fermion  sector.
This LV model is not of Lifshitz type, in the sense that there is isotropic scaling between time and space coordinates, but there is a mass scale to suppress the LV spatial-derivative terms.
The Lorentz-violating Lagrangian considered in \cite{alexandre} reads:
\be\label{bare}
{\cal L}=-\frac{1}{4}F^{\mu\nu}\left(1-\frac{\Delta}{M^2}\right)F_{\mu\nu}
-\frac{\xi}{2}\partial_\mu A^\mu\left(1-\frac{\Delta}{M^2}\right)\partial_\nu A^\nu
+i\ol\psi\br D\psi,
\ee
where $\xi$ is a covariant gauge-fixing parameter (GFP), $D_\mu=\partial_\mu+ieA_\mu$, and $\Delta=\partial_i\partial^i=\vec\partial\cdot\vec\partial$. Our conventions for the
the metric  are (-1, 1, 1, 1).

As emphasized in \cite{alexandre}, no higher order space derivatives are introduced for the fermions, to avoid the introduction of extra non-renormalizable couplings into the theory.
Indeed, in order to respect gauge invariance, which is a crucial assumption of the model, such terms would need to be of the form
\be\label{fermiho}
\frac{1}{M^{n-1}}\ol\psi (i\vec D\cdot\vec\gamma)^n\psi~~~~~~n\ge 2,
\ee
thereby leading to non-renormalizable couplings.

The standard (3+1)-dimensional $QED$ in a covariant gauge is recovered in the limit $M\to\infty$. This scale parametrizes the region of energies in which the Lorentz-violating effects become important. It may or may not be the Planck scale, depending on the microscopic origin of Lorentz Violation.

As discussed in \cite{alexandre}, the Lorentz-violating terms play a dual r\^ole:

\begin{itemize}
\item{(i)} First, to introduce a mass scale, $M$, necessary to generate a fermion mass~\cite{alexandre}:
\be\label{dynmass}
m_{dyn} = M {\rm exp}\left(-\frac{2\pi}{\left(4 + (\xi - 1)\right)\alpha}\right)~,
\ee
where $\alpha = e^2/4\pi$ is the fine structure constant.
Here one draws an analogy with the magnetic catalysis phenomenon of standard QED, studied extensively in the past~\cite{B,gms}, according to which the presence of a sufficiently strong magnetic field catalyzes dynamical generation of a fermion mass for arbitrarily weak QED couplings. This case is also an example of a Lorentz-violating situation: the Lorentz symmetry breaking  is provided by the direction of the external background field. However, there are two important differences from the model of \cite{alexandre}. The magnetic field breaks three-dimensional rotational symmetry, and moreover induces an effective dimensional reduction to two dimensions, as a result of the (1+1)-dimensional form of the fermion propagator of the lowest Landau level, which plays a dominant r\^ole in the strong magnetic field case.

\item{(ii)} Second, the higher-derivative Lorentz-violating terms,
provide an effective regularization of the theory, leading to
finite gap equations~\cite{alexandre}.
An important remark is in order here, to avoid confusion. In this approach, the scale
$M$ is \emph{not} the regulator of the theory (\ref{bare}), since it regularizes loops
with an internal photon line only. It is rather a parameter of the model, which physical quantities, like the dynamically generated mass, will depend upon. As we shall discuss later on, in section \ref{sec:dfoam}, in our stringy quantum-gravity model, which leads to the lagrangian (\ref{bare}) as a low-energy field theory limit, this scale will be expressed in terms of fundamental parameters of the underlying string theory.

\end{itemize}

\section{On Dynamical Mass Generation in LV QED and its Gauge (In)Dependence \label{sec:3}}

In this section we present for completeness a concise brief review of the results of \cite{alexandre} on dynamical mass generation (\ref{dynmass}) in the theory (\ref{bare}). We shall put emphasis on the gauge dependence of the result and propose a resolution to this problem, which will lead us to the gauge-parameter-independent, physically observable, value of the dynamical mass.

From the Lagrangian (\ref{bare}), the bare photon propagator is given by~\cite{alexandre}
\be\label{D}
D_{\mu\nu}^{bare}(\omega,\vec p)=-\frac{i}{1+p^2/M^2}\left( \frac{\eta_{\mu\nu}}{-\omega^2+p^2} +
(\xi -1)\frac{p_\mu p_\nu}{(-\omega^2+p^2)^2}\right) ,
\ee
where $p^0= \omega$ and $p^2=\vec p\cdot\vec p$. We thus observe that, because the pole structure is \emph{not} affected  by the LV terms, the \emph{photon} remains \emph{massless} in this minimally LV model~\cite{alexandre}.

The SD equation for the fermion propagator, used in the derivation of (\ref{dynmass}) is~(\emph{c.f}. for instance \cite{miransky}):
\be\label{SD}
S^{-1}-S_{bare}^{-1}=\int D_{\mu\nu}(e\gamma^\mu) S\,\Gamma^\nu,
\ee
where $\Gamma^\nu$, $S$ and $D_{\mu\nu}$ are respectively the dressed vertex, the dressed fermion propagator and the dressed photon propagator. This equation gives an exact
self-consistent relation between {\it dressed}
$n$-point functions, and thus is \emph{non-perturbative}.
As a consequence, no redefinition of the bare parameters in the theory can be done in order to absorb the would-be divergences, and for this reason one needs this equation to be regularized by the scale $M$, which thus acquires physical significance.

In \cite{alexandre} the so-called \emph{ladder} approximation has been assumed in order to solve the SD equation (\ref{SD}). According to this approximation, corrections to the vertex function, which otherwise would have led to a system of coupled SD equations and would complicate matters significantly, are ignored. It is well-known that this approximation is \emph{not gauge invariant }~\cite{miransky},
as is the case actually with all off-shell field theoretic quantities that are involved at intermediate stages of
calculations of physical quantities in field theory. There are some gauges, termed non-local gauges, in which the bare approximation to the vertex is argued to be an exact ansatz~\cite{gms}. In our discussion below we shall
restrict our analysis to one loop but attempt to determine the gauge-parameter-independent part of the dynamical mass generation.

In this spirit, ref. \cite{alexandre} neglected the loop corrections to the photon propagator, as well as the fermion wave function renormalization, keeping only the corrections to the electron self-energy.
The dressed fermion propagator can then be expressed as (in our conventions):
\be\label{G}
S(\omega,\vec p)=i \,\frac{ p_\mu \gamma^\mu - m_{dyn}}{p_\mu p^\mu + m_{dyn}^2 },
\ee
where $m_{dyn}$ is the fermion dynamical mass.

With these approximations, the SD equation (\ref{SD}), involving a convergent integral,
due to the $M$ dependent Lorentz-violating terms, becomes
\be
m_{dyn}=\frac{\alpha}{\pi^2}\int \frac{d\omega ~p^2dp}{1+p^2/M^2}\frac{m_{dyn}(3+\xi)}{(-\omega^2+p^2)(-\omega^2+p^2+m^2_{dyn})},
\ee
where the fine structure constant is $\alpha=e^2/4\pi$.
This equation has the obvious solution $m_{dyn}=0$, and potentially a second solution, which must
satisfy the following gap equation, obtained after integration over the frequency $\omega$,
\be\label{gap}
\frac{\pi}{(3 +\xi)\alpha}=\int_0^\infty\frac{x\,dx}{1+\mu^2x^2}\left( 1-\frac{x}{\sqrt{1+x^2}}\right),
\ee
where $\mu=m_{dyn}/M$ is the dimensionless dynamical mass, expected to be very small.

Both terms in the last equation, if taken separately, lead to diverging integrals. However, in this specific
combination the divergences cancel each other. As explained in \cite{alexandre},
after some approximations in the limit $\mu \ll 1$, the non-trivial fermion dynamical mass arises as a consistent solution of the equation (\ref{gap}), and is given by (\ref{dynmass}). The selection by the physical system of the dynamical mass $m_{dyn}$ takes place in order to
avoid \emph{Infra Red (IR) instabilities}, which would otherwise occur.

The expression (\ref{dynmass}) for $m_{dyn}$ is not analytic in $\alpha$,
so a perturbative expansion cannot lead to such a result. This
justifies the use of a non-perturbative approach, like the SD in \cite{alexandre}.

There is, however, an obvious drawback of this solution, namely its \emph{dependence} on the \emph{gauge-fixing parameter} $\xi$. The latter  has consequences on the value of $m_{dyn}$, and prompts the question as to what is, if any, the true physical value of the dynamically generated mass.
A possible resolution of this problem, which could lead to the true \emph{physical value} of the dynamical \emph{mass}, may be achieved by the so-called Pinch Technique (PT)~\cite{Cornwall:1982zr} and amounts to computing the dynamical mass in the \emph{Feynman gauge} $\xi = 1$, ignoring in the relevant SD analyses the longitudinal parts of the gauge boson propagator. A brief review of this technique, which is widely used in related problems, not only in particle physics~\cite{binosi2009}, but also in some condensed-matter systems~\cite{mavpapav} (involving energy gap generation of relativistic quasi-particle excitations, similar to the mass generation in  particle physics) is given in the Appendix. We also present there the rather straightforward adaptation of the (one-loop) PT analysis of the Lorentz-Invariant QED to the LV model (\ref{bare}), of interest to us here. This is feasible due to the special (minimal) form of LV in this model.

The consequences of the application of PT for the dynamical mass generation (\ref{dynmass}) in the theory (\ref{bare}) may be summarized in the statement that the physical (\emph{gauge parameter independent}) value of the infrared fermion mass is the one corresponding to the Feynman gauge $\xi = 1$, in which all longitudinal parts of the photon propagator vanish:
\be\label{dynmass2}
m_{dyn} = M {\rm exp}\left(-\frac{\pi}{2 \alpha}\right)~.
\ee
Unfortunately, for standard QED coupling, $\alpha \simeq 1/137$, this value is too small to have any phenomenological significance, even for scales $M$ as high as the Planck scale. One needs enormously transplanckian $M$ in order to get values of $m_{dyn}$ that are near 0.5 MeV,
and thus identify the above model as physical.

One might think of using improved ladder approximations, so as to replace the bare QED coupling $\alpha$ by a running one, which could be stronger and thus increase the
value of the dynamical mass. But it is doubtful that such improved analyses, which at any rate have to be performed for completeness, will yield phenomenologically acceptable values for the electron mass. The only case where such analysis works, in the sense of enhancing significantly dynamical masses, compared to the ladder approximation, is the magnetic catalysis~\cite{gms} in the presence of an external magnetic field, which, as mentioned already, is another example of explicit Lorentz violation due to the direction of the field. However, the reason why in such a case the dynamical mass generation is not so suppressed in the improved approximation is the dimensional reduction to effectively two dimensions induced by the magnetic field. There is no such a reduction in our LV case.

Nevertheless, the use of Lorentz violating theories as seeds for dynamical mass generation is interesting, and may lead to phenomenologically realistic situations if embedded in more complicated theoretical frameworks, like those involving, \emph{e.g.}, hidden sectors in string or other higher-dimensional extensions of the standard model  \`a la Randall-Sundrum scenarios~\cite{RS}, as we now proceed to discuss.

\section{Geometrical Enhancement of Dynamical Mass \label{sec:4}}

In this section we shall present a geometric mechanism for the enhancement of the dynamical mass (\ref{dynmass2}), by embedding the model in higher-dimensional set-ups involving brane worlds in warped bulk geometries, following the spirit of \cite{RS}. In \cite{RS} a large mass hierarchy between Planck masses and TeV scales comes from brane world scenarios in which our world is a negative tension brane, lying at a distance $r_c \pi$ in a five-dimensional bulk from a hidden brane. The five dimensional metrics, which are solutions of Einstein's equations in a (necessarily) Anti-de-Sitter bulk space, have warped factors
\be
ds^2 = e^{-\sigma(z)} \eta_{\mu\nu} dx^\mu dx^\nu + dz^2
\label{warp}
\ee
where $z$ is the bulk (fifth) dimension, and $x^\mu$ are coordinates in our four-dimensional space time.

Because of the warp factor $e^{-\sigma(z)}$ in the metric, fields on our world of mass $m_0$, with canonically normalized kinetic terms, will have physical masses of the form
\be\label{rshier}
m_{phys} = m_0 e^{-\sigma(z_i)}
\ee
where $z_i$ denotes the location of our brane world along the bulk dimension. If, as natural, $m_0$ is of the order of the (reduced) four-dimensional Planck mass $2 \times 10^{18}$~GeV, then a large hierarchy between the Planck scale
and the particle masses $m_{phys}$ in our world may be provided, depending on the size of the exponent $\sigma(z_i)$ of the warp factor.

In the Randall-Sundrum (RS) scenario~\cite{RS}, involving just two branes, with opposite tensions, our world is identified with the negative tension one, located at $z_i = r_c\pi$ and $\sigma = -k |z|$, $k > 0$.
In this scenario, the hierarchy is resolved for $k r_c \pi \simeq \mathcal{O}\left(50\right)$.
Notice that in RS the exponent $\sigma(z_i)$ is positive, and thus the hierarchy factor only decreases the masses as compared to $m_0$.

In \cite{MR}, a more complicated scenario, involving many brane worlds, and higher order curvature terms in the bulk (of Gauss-Bonnet type) has revealed the possibility of exact solutions to the respective bulk Einstein equations, which however allowed for our world to be identified with a positive tension brane, and moreover, and more importantly for our purposes here,  for the exponent of the warp factor to take \emph{negative} values, $\sigma(z_i) < 0$, thus introducing an inverse RS hierarchy.

The bulk geometry is characterized by a gravitational action which includes Gauss-Bonnet higher-derivative corrections:
\bea
S= S_5 + S_4
\label{s5s4}
\eea
where $S_5$:
\bea
 S_5 &=&
\int d^5x \sqrt{-g} \left[ -R -\frac{4}{3}\left(\nabla_\mu \Phi \right)^2
+ f(\Phi) \left(R^2 -4 R_{\mu\nu}^2 +
R_{\mu\nu\rho\sigma}^2\right) \right.\nonumber \\
&~& + \xi(z) e^{\zeta \Phi}
+ \left. c_2~f(\Phi)\left(\nabla _\mu \Phi \right)^4 + \dots \right]
\label{actionGB}
\eea
with $\Phi (z)$ the dilaton field, and
the $\dots$ denoting other types of contraction
of the four-derivative dilaton terms which will not be of
interest to us here, given that by appropriate field redefinitions,
which leave the (perturbative) string amplitudes invariant,
one can always cast such terms in one of the above forms.

The above action is compatible with closed  string amplitude computations
in the five-dimensional space times. Such a compatibility is necessary
in view of the assumption of closed
string propagation in the bulk.
In the stringy case one has~\cite{MR}
$ f(\Phi)=\lambda~e^{\theta\Phi}~,~
\lambda =\alpha '/8g_s^2 > 0~,~c_2=\frac{16}{9}\frac{D-4}{D-2}$,
$\zeta=-\theta=\frac{4}{\sqrt{3(D-2)}}$,  where $\alpha ' = 1/M_s^2$ is the Regge slope, $M_s$ is the string mass scale, $g_s$ is the string coupling and $D (=5)$ is the number of space-time
dimensions.

The four-dimensional part $S_4$ of the action (\ref{s5s4})
is defined as:
\bea
 S_4 = \sum_{i} \int d^4x \sqrt{-g_{(4)}} e^{\omega \Phi} v(z_i)
\label{s4}
\eea
where
\bea
g_{(4)}^{\mu\nu}=\left\{
\begin{array}{l}
g^{\mu\nu}\ , \,\mu,\nu<5\\
0 \ \ \ ,\ \mbox{otherwise}\\
\end{array}\right.
\eea
and the sum over $i$ extends over D-brane walls located at $z=z_i$
along the fifth dimension. Embedding the model (\ref{bare}) in such scenarios, identifies (\ref{bare}) with the effective four-dimensional field theory Lagrangian that describes the low-energy dynamics of open strings (representing the photons) with their ends attached to one of the branes that corresponds to the physical world.

\begin{figure}[ht]
\begin{center}
\includegraphics[width=7cm]{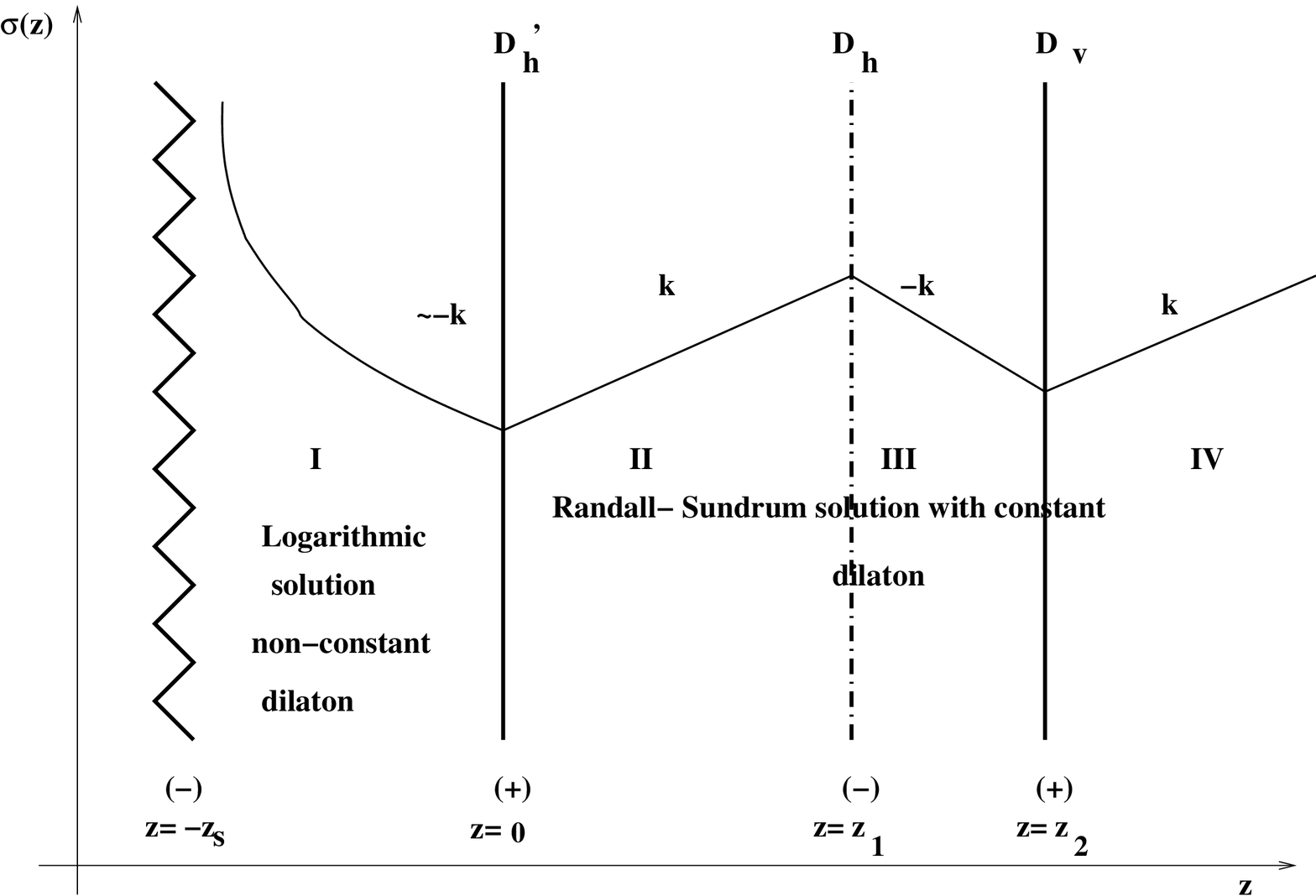} \hfill
\includegraphics[width=7cm]{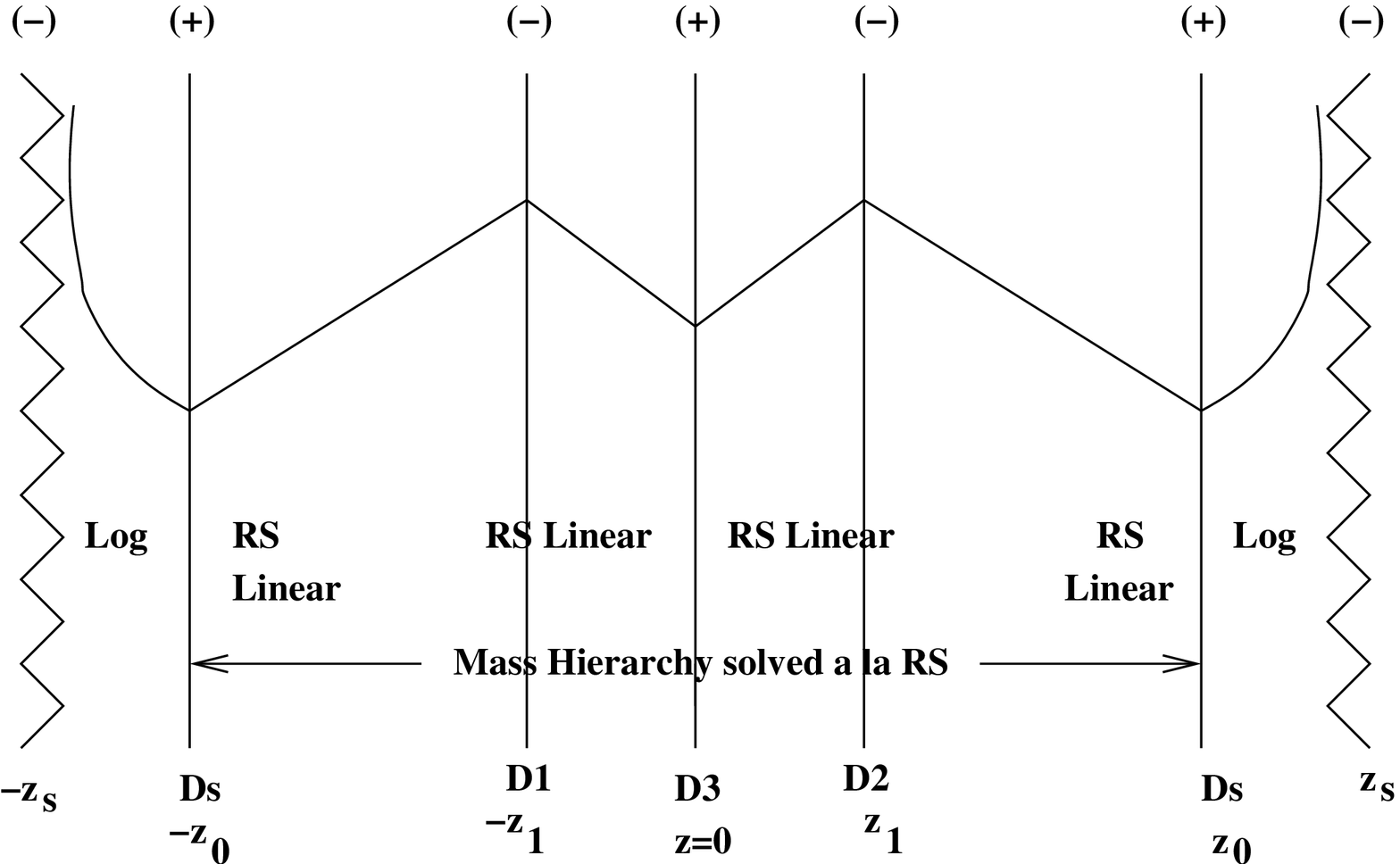}
\caption{\underline{Left figure}:
A multibrane scenario, in which our world is represented
as a positive tension brane (at $z=z_2$), having from the left branes with
alternating-sign tensions, which shield a
bulk naked singularity (which may be thought of
as a limiting (singular) case of a negative tension brane).
To the right of the brane world,
on the other hand,
the bulk dimension extends to infinity. \underline{Right figure}: A multibrane scenario, in which our world is represented
as a positive tension brane (at $z=0$), surrounded by branes
with alternating-sign tensions
that shield two symmetrically-positioned
bulk naked singularities.}
\label{fig:3}
\end{center}
\end{figure}

On assuming warp five-dimensional geometries, of the form (\ref{warp}),
the analysis of \cite{MR}, where we refer the interested reader for details, has demonstrated
that for the actions (\ref{s5s4}), the multibrane situation depicted in fig.~\ref{fig:3}, which involves bulk singularities restricting dynamically the available bulk space, is an exact solution.
In the bulk regions adjacent to the bulk singularities ($z \sim z_s$) one has a logarithmic solution for the warp factor,
$$\sigma (z) = \sigma_2 + \sigma_1 {\rm log}|z -z_s|~,$$
while in the other segments of
the bulk space, induced by the various brane worlds, one has linear solutions, $$\sigma(z) = \sigma_0 + k z~,$$ with the parameter $k$ alternating sign between the various segments of the bulk space, as indicated in fig.~\ref{fig:3}.  Matching the various solutions on each brane, yields a consistent scenario.

The detailed analysis of \cite{MR}, has proven the existence of solutions of the low-energy gravitational bulk equations which imply mass hierarchies of the form:
\be\label{hierarchy2}
       m_{phys} = m_0\,e^{k(2z_1-z_2)}~, \qquad k=\sqrt{\frac{2}{3}}\, g_s M_s > 0~,
\ee
where $z_2$ is the location of our world ($z_2=0$ in the symmetric scenario of the right picture of fig.~\ref{fig:3}). The bulk string scale, $M_s$, is an arbitrary scale in the modern version of string theory and thus can be very different from the four dimensional Planck scale $M_P$.
In this scenario, though, $M_P$ is not lying far from $M_s$, \emph{e.g}. it can be of order~\cite{MR} $M_P \sim M_s/\sqrt{g_s}$. We stress again, that in these scenarios, our physical world is a \emph{positive} tension brane, which makes physical sense.

It is clear that in such a situation, it is possible to have \emph{inverse RS hierarchies}, by arranging appropriately the positions of the various branes. For instance, we may have $z_2 < 2z_1 $.
One may then identify $m_0$ in (\ref{hierarchy2}) with our dynamically generated gauge-invariant mass
(\ref{dynmass2}), in which case the physical mass we would observe in our brane world would be:
\be\label{finalmass}
m_{dyn} = M {\rm exp}\left(-\frac{\pi}{2\alpha} + \sqrt{\frac{2}{3}}\,g_s M_s |z_2 - 2z_1|\right)
\ee
In this way, phenomenologically desired electron masses are obtained by
arranging appropriate the distance $|z_2 - 2z_1|$ in the arrangement of fig.~\ref{fig:3}.

We finally note that, once we puncture the bulk space in such warped scenarios with D-particles, then the quantum-gravity D-foam scenario, discussed in the next section \ref{sec:dfoam}, may be in operation, in which the scale $M$ is related to the string scale $M_s$
and the foam fluctuation parameter via (\ref{MMs}) below.

\section{LV QED as an Effective Theory of a D-particle Space-Time Foam Model\label{sec:dfoam}}

In this section we consider it as useful, or at least intriguing, to attempt and discuss a possible microscopic origin of the model (\ref{bare}) in the context of a stringy space-time foam scenario~\cite{dfoam,westmuckett,emnnewuncert}. We should stress that the material presented here may be omitted by those readers who are not interested in such explanations. Nevertheless, the fact that LV structures, like the ones appearing in (\ref{bare}), characterize parts of the effective action describing the interactions of a space-time foamy defect with photons, gives the model a rather different perspective: it provides, for the first time in our opinion, a concrete realization of the conjecture that space-time foam as a vacuum medium may be responsible for ``slowing down'' particles by giving them a mass. Caution, however, is needed here. The so-generated mass refers to the charged electrons here, the photon remaining massless, as it is revealed by the pole structure of the gauge propagator (\ref{D}) in the model. However, this masslessness should be \emph{disentangled} from an induced non-trivial \emph{vacuum refractive index} in the model, which is associated with purely stringy effects~\cite{emnnewuncert}, namely time-space uncertainties, not captured by the local effective action.

Below we shall demonstrate that LV Lagrangians of the form (\ref{bare}) may arise as (parts of) the low-energy, weak-photon-field limits of an effective Born-Infeld Lagrangian describing the propagation of photons (represented as open strings) on a three spatial brane, punctured by a uniform background of stochastically fluctuating point-like D0-brane defects. Such models have been termed D-particle Foam~\cite{dfoam,westmuckett,emnnewuncert,li} and may find a variety of applications, ranging from providing microscopic string-inspired situations with Lorentz-violating vacuum refractive indices~\cite{emnnewuncert,li,review}, to enhancing dark matter thermal relics, with interesting astro-particle phenomenology~\cite{vergou}.

\subsection{Features of a D-foam model }

\begin{figure}[ht]
\centering
\includegraphics[width=5.5cm]{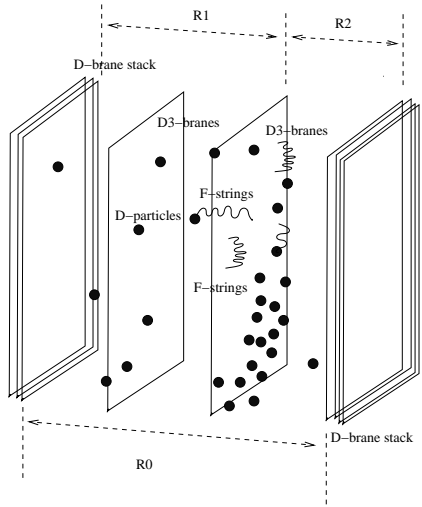}
\caption{Schematic
representation of a generic D-particle space-time foam model. The
model of ref.~\cite{westmuckett}, which acts as a prototype of a D-foam, involves two stacks of D8-branes,
each stack being attached to an orientifold plane. Owing to their special relflective properties, the latter provide
a natural compactification of the bulk dimension. The bulk is punctured by D0-branes (D-particles), which are allowed in the type IA string theory of \cite{westmuckett}. The  presence of a D-brane
is essential due to gauge flux conservation, since an isolated
D-particle cannot exist. Open strings live on the brane world, representing Standard Model Matter and
they can interact in a topologically non-trivial way with the D-particle defects in the foam.}%
\label{fig:dfoam}%
\end{figure}

The basic idea of the D-foam can be described as follows (see, fig.~\ref{fig:dfoam}).
A (possibly compactified from higher-dimensions) three-brane world is moving in a higher dimensional bulk space, punctured by D-particles. Depending on the string theory considered, the latter can be either point-like D0-branes (type IIA strings)~\cite{dfoam,westmuckett,emnnewuncert} or D3-branes wrapped up around appropriate three cycles (type IIB strings~\cite{li}). As the brane world and the D-particles move in the bulk, they cross each other, and thus, for an observer on the D3-brane, the D-particle defects appear as flashing on and off \emph{vacuum structures}.
In this scenarios, ordinary matter and radiation are represented by open strings with their ends attached on the D3-brane. Gravitational degrees of freedom propagate in the bulk.

There are non-trivial interactions of open strings with D-particles, provided there is \emph{no electric charge} flux along the open string excitations, that is, provided the pertinent excitations are electrically neutral~\footnote{For type IIB strings, where the D-particles are not point like, one may have such interactions between electrically charged excitations and the D-particles, however the foam effects on such charged particles are suppressed~\cite{li} compared to those on neutral particles. In our discussion below, therefore, we shall not differentiate between these two cases.}. This is because the electrically neutral D-particles can `cut' an open string, leading to the emission of intermediate strings stretched between the D-particles and the D3 brane~\cite{emnnewuncert,review}. If the open string state carries electric flux, such a cutting procedure
is not allowed, due to charge conservation.

As discussed in \cite{review}, the first-quantization picture for an open string-D-particle interaction, is provided by a world-sheet $\sigma$-model with the following deformation:
\begin{equation}
\mathcal{V}_{\rm{recoil~velocity~part}}^{\rm impulse}=\frac{1}{2\pi\alpha '}
\sum_{i=1}^{D}\int_{\partial D}d\tau\,u_{i}%
X^{0}\Theta\left(  X^{0}\right)  \partial_{n}X^{i}. \label{fullrec}%
\end{equation}
where $M_s$ is the string (mass) scale, $g_s$ is the string coupling, and $u_i$ is the
recoil velocity of the D-particle, the latter assumed heavy (in fact $u^i$ is the spatial part of the
\emph{four-velocity} of the D-particle, but for non relativistic, slow moving, heavy D-particles this is well approximated by the ordinary velocity, to leading order). $D$ in the sum denotes the appropriate  number of spatial target-space dimensions. For a recoiling D-particle confined on a D3 brane, as is our case here, $D=3$.
The operator $\Theta_\varepsilon (X^0) = -i \int_{-\infty}^\infty \frac{d\omega}{\omega + i\varepsilon}$, $\varepsilon \rightarrow 0^+ $, is a regularized Heaviside world-sheet operator.

There is a specific type of conformal algebra, termed logarithmic conformal algebra~\cite{lcft}, that the recoil operators satisfy~\cite{recoil,szabo}.
This algebra is the limiting case of world-sheet algebras that can still be classified by conformal blocks. The impulse operator
$\Theta(X^0)$ is regularized so that the logarithmic conformal field theory algebra is respected~\footnote{This can be done by using the world-sheet scale, $\varepsilon^{-2} \equiv {\rm ln}\left(L/a\right)^2$, with $a$ an Ultra-Violet scale and $L$ the world-sheet area, as a regulator~\cite{recoil,szabo}: $\Theta_\varepsilon (X^0) = -i\,\int_{-\infty}^\infty \frac{d\omega}{\omega- i\varepsilon} e^{i\omega X^0}$. The quantity $\varepsilon \to 0^+$ at the end of the calculations.}.
The conformal algebra is consistent with momentum conservation during recoil~\cite{recoil,szabo}, which allows for the expression of the recoil velocity $u_i$ in terms of momentum transfer during the scattering
\begin{equation}
u_i = g_s\frac{p_1 - p_2}{M_s}~,
\label{recvel}
\end{equation}
with $\frac{M_s}{g_s}$ being the D-particle ``mass'' and $\Delta p \equiv p_1 - p_2$ the associated momentum transfer of a string state during its scattering  with the D-particle.

We next note that one can write the boundary recoil/capture operator (\ref{fullrec}) as a total derivative over the bulk of the world-sheet, by means of the two-dimensional version of Stokes theorem. Omitting from now on the explicit summation over repeated $i$-indices, which is understood to be over the spatial indices of the D3-brane world, we write then:
\begin{eqnarray}\label{stokes}
&& \mathcal{V}_{\rm{recoil~valocity~part}}^{\rm impulse}=\frac{1}{2\pi\alpha '}
\int_{D}d^{2}z\,\epsilon_{\alpha\beta} \partial^\beta
\left(  \left[  u_{i}X^{0}\right]  \Theta_\varepsilon \left(  X^{0}\right)  \partial^{\alpha}X^{i}\right) = \nonumber \\
&& \frac{1}{4\pi\alpha '}\int_{D}d^{2}z\, (2u_{i})\,\epsilon_{\alpha\beta}
 \partial^{\beta
}X^{0} \Bigg[\Theta_\varepsilon \left(X^{0}\right) + X^0 \delta_\varepsilon \left(  X^{0}\right) \Bigg] \partial
^{\alpha}X^{i}
\end{eqnarray}
where $\delta_\varepsilon (X^0)$ is an $\varepsilon$-regularized $\delta$-function.

For relatively large times after the after the moment of impulse, $X^0 = 0$, at which the open string state splits into intermediate open-string ones, as a result of the topologically non-trivial interactions with the D-particle,
which we assume for our phenomenological purposes in this work, the expression (\ref{stokes}) is equivalent to a deformation describing an open string propagating in an antisymmetric  $B_{\mu\nu}$-background ($B$-field) corresponding to a constant external  ``electric'' field in target-space:
\begin{equation}\label{Bfoam}
T^{-1} \, B_{i0} = - T^{-1}\, B_{0i} =  u_i = \frac{g_s \,\Delta p_i}{M_s}~, \qquad T=\frac{1}{2\pi \alpha '}~,
\end{equation}
where $T$ denotes the (open) string tension, $0$ is a temporal index, $i$ is a spatial index. The reader should notice here the phase-space dependence of the background field, which resembles an `electric' field background but in momentum space~\cite{review}, and therefore of Finsler type~\cite{finsler}.

In the above analysis we have ignored a possible angular momentum operator, which also arises as a result of the non-trivial scattering of photons with the D-particle defects. At a $\sigma$-model level, the latter is also described by a logarithmic conformal algebra deformation, which for a three-brane, which we restrict our attention to here, assumes the form:
\begin{eqnarray}
V^{\rm impulse}_{\rm ang~mom~D-part} &=& T^{-1} \, \int_{\partial \Sigma} u^i \epsilon_{ijk} X^j \Theta_\varepsilon (X^0) \, \partial_n X^k \nonumber \\ &=& T^{-1}\,\int_{\rm \Sigma} \varepsilon^{\alpha\beta} \left( u^i \epsilon_{ijk} \Theta_\varepsilon (X^0) \, \partial_\alpha X^j \partial_\beta X^k
+ u^i \epsilon_{ijk} X^j \, \delta_\varepsilon (X^0)\partial_\alpha X^0  \, \partial_\beta X^k \right)\nonumber \\
\label{vertexangmom}
\end{eqnarray}
where we have again applied the two-dimensional Stokes theorem ($\alpha, \beta=1,2$ are world-sheet indices, and $\epsilon^{\alpha\beta}$ is the world-sheet Levi-Civita tensor). On the other hand, $\epsilon_{ijk}$ is the antisymmetric symbol in the three spatial dimensions of the brane world.
 The logarithmic conformal properties~\cite{recoil} of the deformation arise from the $X^J$ parts.  For relatively large times $X^0 > 0$ after the impulse, we consider here, we may ignore the $\delta$-function terms, and in this case the effects of the angular momentum deformation in target space are equivalent to the open string propagating in an antisymmetric tensor ``magnetic-field'' type background with spatial components given by
\begin{equation}\label{magfield}
T^{-1} B_{ij} = - T^{-1} B_{ji} = \epsilon_{ijk} u^k~.
\end{equation}
This should be combined with (\ref{Bfoam}) in order to provide a complete description of the \emph{average} interactions of the photons with the D-foam, in a first quantized version.

\subsection{Finsler-Born-Infeld (FBI) Effective Actions}

The effective target-space action on the D3-brane world, where the D-particle meets the open string photon state,
is described by the following Born-Infeld Lagrangian~\cite{seibergwitten,sussk1}:
\begin{equation} \label{bi}
S_{BI} = \frac{T^2}{g_s} \int d^4 x \sqrt{{\rm det}\left(g + T^{-1} (B + F)\right)}~.
\end{equation}
where $F_{\mu\nu}=\partial_\mu A_\nu - \partial_\nu A_\mu$ is the Maxwell tensor of the photon field $A_\mu$.

In fact, the field theory of photons in the presence of the `electric' field type background $B_{0i}$ implies a time-space non-commutativity for the model~\cite{review}, with non commutativity parameter $u_i$ which is of Finsler type~\cite{finsler}, in the sense that it is proportional to the photon momentum transfer (\ref{Bfoam}).
On the other hand, the presence of magnetic field type backgrounds (\ref{magfield}) also imply Finsler-type spatial non commutativity among $X^j$ target space coordinates.

As a result of this, it is not in general possible to write down a local effective action for the description of the photon-D-particle interactions. The Lagrangian (\ref{bi}) depends on both space-time coordinates and the \emph{momentum transfer}, which cannot be expressed as an ordinary local term in an effective action framework.
Nevertheless, as we shall argue now, there is an approximation, namely the \emph{stochastic D-foam} background, for which construction of a low-energy \emph{local} effective action becomes possible and, in fact, for weak photon fields, and under some other special circumstances to be described below, assumes the Lorentz-violating form (\ref{bare}).

To this end, we first remark that, in the phase space of a D3-brane world, the function $u_{i}$, (\emph{c.f.} (\ref{Bfoam})), involving a momentum transfer, $\Delta p_{i}$, can be \emph{modelled} by
a local operator using the following parametrization~\cite{sarkar}:
\begin{equation}
u_{i}=g_{s}\frac{\Delta p_{i}}{M_{s}}=\frac{g_s}{M_s} r_{i}p_{i}\,~,\, \quad {\rm no~sum~over}\, \quad i=1,2,3~,\label{defu2}\end{equation}
where the (dimensionless) variables $r_{i},i=1,2,3$, appearing above,
are related to the fraction of momentum that is transferred at a collision
with a D-particle in each spatial direction $i$. In the \emph{stochastic foam approximation}~\cite{sarkar}, these parameters are taken as Gaussian normal random variables with a range $-\infty$
to $+\infty$ and defining moments \begin{equation}
<r_{i}>=0,\label{moment1}\end{equation}
 \begin{equation}
<r_{i}r_{j}>=0, \quad \textrm{if}\, i\neq j\label{moment2}\end{equation}
 and \begin{equation}
\sigma_{i}^{2}=<r_{i}^{2}>-<r_{i}>^{2}=<r_{i}^{2}>\neq0.\label{moment3}\end{equation}
 An isotropic foam situation, which we consider here, would require $r_{i}=r$, for all $i=1,2,3$.
In such a case the variances
\begin{equation}\label{varianceisotr}
<(r_{i})^{2}>=\sigma^{2}~, \qquad i=1,2,3~,
\end{equation} are equal along all
spatial directions.

However, for fully rotational invariance in three space of the Born-Infeld action (\ref{bi}),
involving interactions between the velocity fields $u_i$ and the photon Maxwell tensor $F^{\mu\nu}$,
one more requirement is necessary. When considering
the application of the above averages to the recoil velocities, the latter must assume the form:
\bea\label{gaussian2}
<u_i > = 0~, \qquad
  < u_i u_j > = \delta_{ij} \frac{\sigma^2\,g_s^2}{M_s^2} p^k p_k
\eea
The averages $< \dots >$ denote \emph{both} statistical averages, over populations of D-particle defects in the foam (\emph{c.f}. figure fig.~\ref{fig:dfoam}), as well as target-space quantum fluctuations. The latter can be induced by considering appropriate summation over world-sheet genera~\cite{szabo}.
This is quite important for our purposes and we shall come back to this issue and its implications later.

For the moment, we remark that the stochasticity conditions (\ref{moment1})  imply \emph{restoration} of Lorentz-invariance as an average phenomenon, with non-trivial fluctuations expressed in the isotropic and three-space rotationally invariant case by (\ref{moment2}), (\ref{varianceisotr}) and (\ref{gaussian2}). Hence, although at individual scatterings of photons with D-particles, Lorentz invariance (and isotropy of space) will be lost locally, due to the presence of the recoil velocity of the D-particle, $u_i$, the isotropic foam washes out on average such violations and isotropy and rotational invariance are restored.

Let us now go back to the effective action (\ref{bi}). On defining the generalized field $\mathcal{F}_{\mu\nu} \equiv F_{\mu\nu} + B_{\mu\nu}~$, we make use of the fact~\cite{tseytlin}
that in four space-time dimensions (assumed Minkowski flat for concreteness from now on), the determinant
${\rm det}_4 \left(\eta + T^{-1}\,F\right)$ has special properties, that allow the following representation of the Born Infeld action (we work from now on in units where the string tension is $T=1$)~\footnote{The reader is reminded at this stage that the indices are raised and lowered with the background metric $g$, which in our case, has been assumed to be Minkowski flat, for brevity and definiteness. In a general situation, where the metric $g$ is not trivial, the pure foam contributions, proportional to the various powers of $u_i^2$ contribute to (dark) vacuum energy~\cite{westmuckett,emnnewuncert} and can be constrained by cosmological considerations, for instance. We ignore such terms for our discussion here.}:
\bea\label{4dbi}
S_{BI} & = &  \frac{1}{g_s} \, \int d^4 x \left(I_2 + I_4 \left(1 + \mathcal{O}(\mathcal{F}^2) \right)\right) +{\rm const}
\nonumber \\
\quad I_2  &=& \frac{1}{4}\mathcal{F}_{\mu\nu}\mathcal{F}^{\mu\nu}~, \quad I_4 = -\frac{1}{8} \left[\mathcal{F}_{\mu\nu}\,\mathcal{F}^{\nu\rho} \, \mathcal{F}_{\rho\lambda}\,\mathcal{F}^{\lambda\mu} -
\frac{1}{4}\left(\mathcal{F}_{\mu\nu}\mathcal{F}^{\mu\nu}\right)^2\right]~.
\eea
In the weak photon field approximation, of interest to us here, we shall ignore terms of order higher than quadratic in the photon field and the (small) recoil velocity $u_i$ field. This is a consistent approximation for relatively heavy D-particles, whose recoil is suppressed by their mass.
We also take a quantum average over stochastic fluctuations of the $B$-field, using (\ref{gaussian2}), and keep terms quadratic in photon ($A$) or (averaged) recoil ($<u^2>$) fields, including mixed terms of order $A^2 <u^2>$.

\subsection{Target-Space Quantization Proposal for the FBI Action and Minimal LV QED}

There is an important and novel feature associated with the proper \emph{quantization} in target space of the Finsler background $B_{0i}$ (\ref{Bfoam}), which is provided by summation over world-sheet genera. As discussed
in detail in ref.~\cite{szabo}, making proper use of the (logarithmic) conformal field theory properties of the D-particle recoil vertex operator in a $\sigma$-model approach~\cite{recoil}, the summation over world-sheet topologies results in quantum uncertainty relations involving the fluctuations of the recoil velocities and of the collective coordinates describing the initial position of the D-particle. Such uncertainties have been shown to correspond to those induced by canonically quantized collective D-particle momentum and position operators in target space. In this way, it has been argued in \cite{szabo} that summation over world-sheet genera converts,
 via (\ref{defu2}), the recoil velocity fields $u_i$ into quantum fluctuating momentum operators in target space~\footnote{Technically speaking, we should mention here that the effects of the summation over genera on the fluctuations of the background fields $u_i$ can be expressed in closed form only in the bosonic string case. In the world-sheet supersymmetric case~\cite{szabosusy}, such closed expressions were not possible. Nevertheless, this does not invalidate our arguments on the correspondence principle (\ref{quant}), which we conjecture to characterize all types of string models.}.

More specifically, one has the correspondence:
\bea\label{quant}
B_{i0} = u_i  \Rightarrow  \widehat{B}_{i0} = \widehat{u}_i = -i g_s \, \frac{r_i}{M_s} \hbar \frac{\partial}{\partial X^i} &\equiv & -i g_s \, \frac{r_i}{M_s} \hbar \nabla_i~,
\quad  {\rm no~sum~over~} i = 1,2,3 ~. \nonumber \\
 B_{ij} = \epsilon_{ijk}\,u^k & \Rightarrow &  \widehat{B}_{ij} = \epsilon_{ijk}\,\widehat{u}^k~.
\eea
In view of this, the non commutativity uncertainty relations $[ X^i, t ] \sim u^i$ of the classical recoil background~\cite{review} become now complicated quantum operator relations. En passant, we also note that the correspondence (\ref{quant}) can also lead to master equations for the study of the induced decoherence of quantum matter propagating in the stochastic D-foam quantum-fluctuating backgrounds~\cite{sarkar}.

We now wish to make a technical but important observation regarding the correspondence (\ref{quant}), which we shall make use of in our analysis below.
As already mentioned, we have inferred this correspondence by studying the corresponding $\sigma$-model approach to recoil~\cite{recoil}, in which, in order to guarantee the convergence of the world-sheet path integrals, one is forced to use time fields $X^0$ with \emph{Euclidean} signature. In this sense, the correspondence (\ref{quant}), should be augmented to
include the following:
\bea\label{eucl}
&& B^{i0} \Rightarrow \widehat{B}^{i0} =   g^{00}_E g^{ik}_E\widehat{B}_{k0} = + \widehat{B}_{i0}~,
\eea
where the subscript $E$ indicate \emph{Euclidean signature}. Only at the very end of the computations, after we replace the background $B$-field by appropriate operators,
and only then, we revert back to Minkowski signature, by means of \emph{analytic continuation}.

In this picture, where the quantum fluctuating aspects of the recoil operators are taken care of by means of the correspondence (\ref{quant}), (\ref{eucl}), the statistical aspects of the foam, are implemented by averaging
the momentum transfer statistical variable $r$ ($<...>$) over populations of quantum fluctuating
D-particles, using   the relations (\ref{gaussian2}) for the case of isotropic foam.
In fact, we now have the correspondence:
\bea\label{gaussian}
<u_i > = 0~, \qquad
  < u_i u_j > = \delta_{ij}\, \frac{\sigma^2\,g_s^2}{M_s^2} p^k p_k \quad \Rightarrow \quad - \hbar^2 \frac{\sigma^2\, g_s^2}{M_s^2} \delta_{ij} \Delta~, \quad \Delta \equiv {\vec \nabla} \cdot {\vec \nabla}~.
\eea
This is our \emph{prescription}, which, as we shall show below, maps the effective action (\ref{4dbi}) to  particularly simple local effective actions containing terms of the form (\ref{bare}).

Indeed, on implementing (\ref{quant}), (\ref{eucl}), the relevant photon-dependent terms of the Finsler-Born-Infeld (FBI) action, which we restrict our attention upon in this work, assume the form
\begin{eqnarray} \label{bioper}
&&S_{BI} \ni \frac{1}{g_s} \,\int d^4 x  : \left[\frac{1}{4} F_{\mu\nu}\left(1 +
\frac{b}{16}\,\widehat{B}_{\alpha\beta}\widehat{B}^{\alpha\beta}\right)F^{\mu\nu}
+ \frac{a}{64} F_{\mu\nu}\,\widehat{B}^{\mu\nu}\widehat{B}_{\alpha\beta}\,F^{\alpha\beta} \right.
\nonumber \\
&&\left. -\frac{1}{8}\,F_{\mu\nu} \widehat{B}^{\nu\rho} \widehat{B}_{\rho\beta} F^{\beta\mu}\right] : + \dots
\end{eqnarray}
The $\dots$ represent higher order terms in the fields $B$ and $F$ and the symbol : \dots : denotes \emph{appropriate quantum operator ordering} and the ordering constants $a, b$ are such that
\be\label{ordering}
a + b =2~.
\ee
 Above we have ignored terms involving odd powers of the operators $\widehat{B}_{\mu\nu}$ since such terms vanish in our stochastic Gaussian background (\ref{gaussian}). Also we dropped terms involving the operators $\widehat{B}_{0i}$ or $\widehat{B}_{ij}$ lying on the far left-hand-side of the integrand in (\ref{bioper}), as these correspond to total spatial derivative terms
and hence do not contribute, upon the assumption that the fields decay away at spatial infinity on the brane world.

Upon recalling that, when making the correspondence (\ref{gaussian}), one uses as an intermediate step target-space times with Euclidean signature, \emph{i.e.} (\ref{eucl}), we observe, after some straightforward algebra, that (\ref{bioper}) reduces to (using (\ref{ordering})):
\begin{eqnarray}\label{bioperfinal}
S_{BI} \ni \frac{1}{g_s} \,\int d^4 x   \left[\frac{1}{4} F_{\mu\nu}\left(1
 + \frac{1}{4}\,(1 - \frac{b}{2})\, \frac{g_s^2 \sigma^2}{\,M_s^2} \Delta \right) \,F^{\mu\nu} \right] \, + \, \dots ~,
\end{eqnarray}
where the $\dots$ denote terms higher order in derivatives and the Maxwell tensor $F_{\mu\nu}$, and $\Delta$ is the 3-space Laplacian, $\Delta \equiv \nabla_i \nabla^i = \vec{\nabla} \cdot \vec{\nabla}$.

In the classical limit the FBI action is recovered trivially.
The quantum ordering ambiguities is an issue here. Usually, quantum ordering of operators is associated with
hermiticity of the effective lagrangian, which is not an issue here, due to the foam-background stochasticity (\ref{gaussian}). The physics which selects the correct quantum ordering in this picture is encoded in the full (still elusive) underlying theory of quantum gravity in this context, and at this stage the so-constructed effective action should be considered as somewhat phenomenological. For instance, the choice $b=2$ would eliminate any LV terms, however, from the point of view of avoiding infrared (IR) divergences and the associated instabilities, the solution that leads to dynamical mass generation is preferred. Adopting the point of view that the full quantum gravity theory should act as an IR regulator, one then is forced to select an ordering with $b \ne 2$.

Moreover, we observe that any ordering with $b < 2$ would lead to terms which would affect the three-space pole structure of the photon propagator.  Indeed, the photon propagator stemming from (\ref{bioperfinal}), assumes the form (up to gauge fixing terms that we do not write explicitly here):
the bare photon propagator is given by~\cite{alexandre}
\be\label{D2}
D_{\mu\nu}^{bare}(\omega,\vec p)=-\frac{i}{1  - p^2/M^2}\left( \frac{\eta_{\mu\nu}}{-\omega^2+p^2} + {\rm gauge~fixing} \right) , ~\quad  M^2 \equiv \frac{4\,M_s^2}{(1-b/2)\,g_s^2 \, \sigma^2} > 0~.
\ee
Thus, in Fourier space, this would imply that the effective action would no longer be unitary for momenta above the  scale $M$, since the overall signature of the photon propagator would change. This would be fine for a low-energy effective action in a classical background, and indeed this is exactly what happens~\cite{emnnewuncert,review} when the classical recoil velocity
(\ref{recvel}) exceeds the speed of light \emph{in vacuo}. Such mass scales define the range of validity of the low-energy local effective action.

However, in our prescription we would like to go beyond such classical effective field theories. If we could find a quantum ordering which allows extension of the action beyond this cut-off then, in principle, we could describe some aspects of the foam on radiation for arbitrarily
large momenta. This would provide a sort of partial ultraviolet (UV) completion of the low-energy theory (as far as dynamical mass generation is concerned - the reader's attention is called at this stage to the fact that other aspects of the foam, such as vacuum refraction induced photon delays, cannot be described within the framework of local effective field theories~\cite{emnnewuncert,review}, see below). Dynamical mass generation is, of course, an IR phenomenon and one would have thought that the detailed structure of the theory at the UV would not affect it. This is, for instance, what arguably may happen with the Landau pole of QED, whose presence may not affect dynamical mass generation (there is support towards this result from lattice calculations but, as far as we are aware of, no rigorous proof exists as yet~\cite{gms,lpole}). However, this is not quite the case if an effective theory itself breaks down at a given momentum scale, in the sense, as is the case here, that the unitarity of the effective Lagrangian (sign of photon propagator terms) is lost above such scales. If one were to study dynamical generation in such cases, the momentum integrals in the SD  equations had to be cut-off from above at the specific scale $M$. This would be an unwelcome feature for a consistent SD treatment, such as the one in \cite{alexandre}, which requires cancellation of potential UV divergences, and thus implicit extension of the model beyond any UV cut-off.

Fortunately in our model, such UV cut-offs can be avoided by a judicious choice of quantum ordering.
A minimal class of such orderings, which respect the pole structure of the photon propagator (\ref{D2}) in three space, is the one in which the ordering parameter is in the range:
\be
b > 2
\label{physord}
\ee
After a formal analytic continuation back to Minkowski space-time, the corresponding actions (\ref{bioperfinal}) become \emph{precisely} of the form (\ref{bare})
of ref.~\cite{alexandre} with the mass scale:
\begin{equation}\label{MMs}
M = \frac{M_s}{\,g_s \,\sqrt{{\tilde \sigma}^2}}~, \qquad \tilde{\sigma}^2 \equiv \sigma^2 \frac{1}{4}|1 - \frac{b}{2}|~.
\end{equation}
Any residual ordering ambiguity is thus absorbed in the fluctuations of the foam, which is thus a (small) phenomenological parameter in our first-quantized approach. It is hoped that when a full, second-quantized quantum gravity model for D-foam fluctuations becomes available, such ambiguities will be removed~\footnote{We also note that for the unique (in this range of $b$) value $b=10$, $\tilde{\sigma}^2 = \sigma^2 $ and thus the scale $M$ of the LV terms in the action (\ref{bioperfinal}) becomes identical to the one
at which the foam-averaged recoil velocity equals the speed of light in vacuo~\cite{emnnewuncert,review}. This choice of ordering then allows for a unique quantum gravity scale to enter in the model in various forms.}. In this sense the so-chosen quantum ordering leads to an effective action that is characterized by \emph{maximal suppression} of the LV effects, and through mass generation, cures IR instabilities.

We next come to the fermion sector. In view of the properties of the D-foam, according to which the latter is \emph{transparent} to \emph{charged} fermion \emph{fields}, due to charge conservation, the lowest order (in weak field) effective action term in the fermion sector will be given by the \emph{ordinary} QED fermion-photon coupling. There will be no tree-level coupling of fermions to the foam recoil velocity field $u_i$:
\begin{equation}\label{fermion}
S_{\psi} = \int d^4 x {\overline \psi} \gamma^\mu i\,D_\mu \psi~, \quad D_\mu = \partial_\mu + i  A_\mu~.
\end{equation}
Thus, the \emph{renormalizability} argument of \cite{alexandre} on the absence of higher-derivative non covariant fermion-gauge-boson couplings (\ref{fermiho}), is \emph{replaced} here by the \emph{transparency of the D-foam} to charged fermions, as a consequence of charge conservation~\footnote{This argument is strictly speaking valid only for type IIA string theory D-foam, in which the D-particles are point-like. For type IIB string theory D-foam models, on the other hand, D-particles are compactified D3-branes, and as such there are non trivial, but much more \emph{suppressed} fermion-foam couplings~\cite{li}.}.

In view of the dynamical mass generation arguments, presented previously, one expects the D-foam models to lead to masses for fermions of the type (\ref{dynmass}) (or (\ref{dynmass2})), with the scale $M$ being replaced by (\ref{MMs}). Since, in the context of our string-inspired foam and the Born-Infeld action, the gauge coupling is directly related to the string coupling $\sqrt{g_s}$, the dynamical mass (\ref{dynmass}) (or (\ref{dynmass2})) in this case is \emph{non analytic} in \emph{both} the (weak) string coupling and the (small) D-foam fluctuation parameter ${\tilde \sigma}^2$.

The effective action (\ref{bioperfinal}), (\ref{fermion}) describes part of the effects of foam on radiation and charged matter. However, as already discussed in the relevant literature~\cite{review,emnnewuncert}, the induced vacuum refraction cannot be captured solely by this local effective action. The \emph{causal} time delays of the re-emitted photons after their topologically non-trivial interactions with the D-foam, which scale linearly with the photon energy, and are thus suppressed only by a single power of the string mass scale, are purely stringy effects, associated with time-space uncertainties~\cite{sussk1,emnnewuncert,li,review} as a result of intermediate string (and hence non local) states, stretched between the D-particles and the D3 brane universes. On the other hand, the modifications due to the foam in the local string effective action (\ref{bioperfinal}) are quadratically suppressed by the string scale, (\ref{MMs}), as a result of the stochasticity assumption (\ref{gaussian}).

Finally, before closing this section, we mention that by embedding the D-foam model into the multibrane world scenarios of section \ref{sec:4}, we may enhance the dynamical mass generation to phenomenologically acceptable values, via the (inverse) Randall-Sundrum hierarchies (\ref{hierarchy2}), (\ref{finalmass}).

\section{Conclusions and Outlook \label{sec:5} }

In this note I have elaborated on a recently proposed model~\cite{alexandre} of Quantum Electrodynamics with minimal Lorentz Violation, associated with higher-order spatial derivative terms in the Lagrangian, suppressed by a scale $M$, which lead to dynamical mass generation for the charged fermions. I have proposed a way of extracting the physical part of the fermion mass function, which is independent of the gauge-fixing parameter. My arguments were based on the so-called Pinch Technique of Particle Physics. According to this approach, only a certain class of Feynman diagrams can be resummed and play a r\^ole in the evaluation of the final physical
quantities, such as the observable value of the dynamical fermion mass. The upshot of the analysis was that the physical value of the dynamically generated mass corresponds to the one derived in the Feynman gauge, $\xi = 1$. For the problem at hand, this, unfortunately, yields phenomenologically unrealistic small masses for the fermions.

However, if the model is embedded in a multibrane world scenario, with five-dimensional warp bulk geometries, then an enhancement mechanism is possible, through the dependence of the masses on the non-trivial warp factor in the observable (positive tension) brane world. The enhancement presupposes a reversed Randall-Sundrum hierarchy, which is possible in five dimensional models with higher curvature gravitational corrections, of Gauss-Bonnet type (the latter is consistent with the effective target-space-time action corresponding to string theory scattering amplitudes in the bulk space).

 I have also discussed a possible microscopic origin of these LV terms in the context of stringy space-time foam models, in which a brane world, over which photons propagate, is punctured by D-particle defects, with which the photons interact non trivially. Electric charge conservation arguments necessitate that tree level couplings between the foam and the fermions vanish, thereby making the effects of the D-particle quantum fluctuations on the fermions felt only through their coupling with the gauge (photon) fields. On the other hand, the interactions of the photons with the D-particles can be partly summarized by means of a Born-Infeld type effective lagrangian. The latter involves,
in addition to the Maxwell field strength of the photon field, also LV B-type fields, representing the recoil velocity fields of the D-particle defect during its scattering with the photon. The B-fields, become quantum derivative momentum operators upon summing up world-sheet genera, and thus a true quantum D-foam situation is described by a complicated Finsler-Born-Infeld (FBI) lagrangian, which is, in general, non-local.

However, for weak fields and foam, appropriate truncations and approximations can be made, with the implication  that the relevant parts of the effective action, describing the interactions of photons and electrons with the foam, are provided by a (local) effective lagrangian of the form (\ref{bare}), with minimal LV. This exercise had  provided the first concrete realization of the conjecture that the stochastic D-foam is responsible for generating masses for the fermions, by catalyzing, through its Lorentz-violating terms in the gauge sector,
weak-gauge-coupling dynamical fermion mass generation. The reader should bear in mind, though, that this local effective action is \emph{not} capable of capturing the vacuum refraction aspects of the D-foam. The latter are associated~\cite{emnnewuncert} with purely stringy (and thus non-local) time-space uncertainty effects, associated with intermediate string states of finite length, stretched between the D-particle and the brane world, during the quantum scattering of photons off the foamy defects.

Before closing I would like to make some general remarks about effects of the space time foam on neutral fermions, such as neutrinos. In our mechanism,
dynamical mass is \emph{catalyzed} by the LV terms in the gauge action, even for weak gauge couplings, and proceeds  through the coupling of the gauge fields to the \emph{charged} fermions.
In this sense, our space-time foam mechanism does not apply to neutrino (assumed electrically neutral), at least to leading order. However, the neutrinos of the Standard Model eventually couple to photons via effective higher-loop vertices involving the weak-interaction gauge Bosons. In this way one obtains an ``effective running neutrino charge'' and an associated ``radius''~\cite{giundi}. In this sense, the D-foam, through its LV coupling to photons, may also catalyze a (much more suppressed) neutrino mass, and in fact the electric neutrality of the neutrino, would provide a natural reason for the observed hierarchy of the neutrino masses, as compared to the electron mass.
This is something one should look at more carefully. Moreover, since the stochastic effects of the D-foam will violate CPT~\cite{sarkarbeny}, the so-generated neutrino masses may experience a medium-induced CPT violation~\cite{bar}. It would be interesting also to see whether a mass hierarchy among different flavours can be generated this way, for example by means of a judicious combination of LV terms in the gauge sector of the standard model and flavoured neutrino interactions.

We hope to come back to such fundamental issues in future works. For the moment, we can only provide the reader with the above-described toy ideas and speculations on how Lorentz Violating effects in the standard model sector, that may have a microscopic origin in some foamy structures of quantum space time, may be responsible for the generation of fermion masses, through the coupling of the foam ``medium'' with the gauge fields that, in turn, interact with matter fermions. In the toy model we have discussed in this work, the photon gauge field remains massless, as evidenced by the structure of its propagator and one-loop vacuum polarization. In general, this may not be true in more complicated situations of anisotropic foam.

\section*{Acknowledgements}

I wish to thank Jean Alexandre, Domenec Espriu and Joannis Papavassiliou for discussions.

\section*{Appendix : The Pinch Technique (PT) and Gauge-Fixing-Parameter Independent Mass}

\subsection*{General Remarks on PT}

It is well known that off-shell Green's functions
depend in general on
the  gauge-fixing  procedure  used  to  quantize the  theory,  and  in
particular on  the gauge-fixing parameter (GFP) chosen  within a given
scheme. The   fermion  self-energy  $\Sigma  (p)$, for example, of interest to us here, is
GFP-dependent already  at the one-loop  level.
The dependence  on the
GFP is  in general non-trivial and  affects the properties  of a given
Green's  function.   It is understood that,  when  forming
observables,  the gauge  dependencies  of the  Green's functions  cancel
among  each  other order  by  order  in  perturbation theory,  due  to
powerful  field-theoretical  properties. Nevertheless, these dependencies  pose a major difficulty when one  attempts  to   extract  physically  meaningful  information  from
individual  Green's functions.   This  is  the case when studying
the SD equations; this infinite
system of  coupled non-linear integral
equations  for all Green's  functions of
the  theory is inherently  non-perturbative  and can
accommodate phenomena  such as  chiral symmetry breaking  and dynamical
mass  generation.

The main  problem in this context is  that the SD
equations are  built out  of gauge-dependent Green's  functions; since
the  cancellation  mechanism  is  very subtle,  involving  a  delicate
conspiracy of terms  from {\it all orders}, a  casual truncation often
gives   rise   to   gauge-dependent  approximations  for  ostensibly
gauge-independent quantities \cite{Cornwall:1974vz}.
The study of SD
equations, and especially of ``gap equations'',
has been  particularly popular in many studies.

To address the problems of the gauge-dependence of
off-shell Green's functions a method known as the \emph{Pinch Technique} (PT)
has been introduced \cite{Cornwall:1982zr}.
For a detailed and up-to-date review on this technique and its diverse applications, the reader is referred to \cite{binosi2009}. For a comprehensive review, of relevance to our (one-loop approximate) discussion here, and its application to the infrared structure on low-dimensional gauge theories, with potential application to condensed matter systems, such as high-temperature
superconductors, see \cite{mavpapav}.

The PT is  a
diagrammatic  method which exploits  the  underlying  symmetries
encoded  in  a  {\it  physical} amplitude  such  as  an  $S$-matrix  element,
or a Wilson loop,
in  order  to  construct effective   Green's   functions    with   special
properties.    The aforementioned symmetries,  even though they are  always
present, they are  usually concealed by  the gauge-fixing  procedure.
The  PT
makes them  manifest by means  of a  fixed algorithm,  which does  {\it not}
depend on  the gauge-fixing scheme  one uses in order  to quantize the theory,
{\it i.e.} regardless of the  set of Feynman rules used when writing down  the
$S$-matrix  element.   The method  exploits  the elementary  Ward
identities
triggered  by the  \emph{longitudinal  momenta} appearing  inside  Feynman  diagrams
in  order  to  enforce massive cancellations.
The realization of these
cancellations  mixes non-trivially contributions stemming from diagrams of
different  kinematic nature (propagators, vertices, boxes). Thus, a  given
physical  amplitude  is reorganized into sub-amplitudes,   which  have   the
same   kinematic   properties  as conventional $n$-point functions and, in
addition, are  endowed with  desirable physical  properties, such as
GFP-independence.

Finally, the PT amounts to a non-trivial
{\it reorganization of the perturbative expansion}.
The r\^ole of the PT  when dealing with SD equations is to
(eventually)  trade   the  conventional  SD   series  for
another,  written  in terms  of  the  new, gauge-independent  building
blocks \cite{Cornwall:1982zr,Mavromatos:1999jf,Sauli:2002tk}.
The upshot  of this program would then
be  to truncate this  new series,  by keeping  only a  few terms  in a
``dressed-loop'' expansion, and maintain exact gauge-invariance, while
at the same time accommodating non-perturbative effects.
We hasten to emphasize that the aforementioned program is still {\it not}
complete; however, a great deal of important insight
on the precise
GFP-cancellation mechanism has been accumulated, and the
field-theoretic properties of gauge-independent Green's
functions have been established in detail.
The generalization of the PT to all orders for Lorentz invariant quantum gauge field theories has been
recently accomplished \cite{Binosi:2002ft}.

It would be interesting to discuss extensions of these ideas to LV gauge theories, either of Lifshitz type or
in the framework of the Standard Model Extension~\cite{kostel}. In the model (\ref{bare}), however,
the validity of the PT is straightforward, due to the specific minimal form of LV, as we now proceed to discuss.

\subsection*{A One-Loop Application of PT to Dynamical Mass Generation in LV QED}

In this part of the Appendix we explain how the PT
gives  rise to
effective,  gauge-independent  fermion  self-energies at one-loop,
in our Lorentz-Violating QED Model. The discussion, parallels the cases of Lorentz invariant
theories of  QED \cite{Binosi:2002ft,mavpapav}, due to the specific form of the (gauge fixed) photon propagator
(\ref{D}), which we may write compactly as:
\begin{equation}
\Delta_{\mu\nu}(\ell,\xi ) = \left(\Delta _{LV}(\vec{\ell}\cdot\vec{\ell})\right)
\left(-\frac{\D i}{\D \ell^2}
\left[\ g_{\mu\nu} - (1-\xi) \frac{\D \ell_\mu
\ell_\nu}{\D \ell^2}\right]\right)~, \quad \Delta_{LV}(\vec{\ell}\cdot\vec{\ell}) = \frac{1}{1 + \frac{\vec{\ell}\cdot\vec{\ell}}{M^2}}~.
\end{equation}
The off-shell propagator depends of course on the gauge-fixing parameter (GFP) $\xi$. It is important for our discussion below that the Lorentz-Violating (LV) terms
$\Delta_{LV}(\vec{\ell}\cdot\vec{\ell})$, factorize and do not introduce any extra pole singularities in momentum space. The quantity $\ell^\mu$ denotes the contra-variant four momentum $\ell^\mu = (\omega, \vec{p})$.

The form of
$\Delta_{\mu\nu}(\ell,\xi )$ for the special choice   $\xi =1$ (Feynman gauge)
will be of central importance for the PT application in our discussion below. We denote it by
$\Delta_{\mu\nu}^{F}(\ell)$, {\it i.e.}
\begin{equation}
\Delta_{\mu\nu}(\ell, 1 )\equiv \Delta_{\mu\nu}^{F}(\ell)
=  \left(-\frac{\D i}{\D \ell^2}\, g_{\mu\nu}\right)\,  \left(\Delta _{LV}(\vec{\ell}\cdot\vec{\ell})\right).
\end{equation}
$\Delta_{\mu\nu}(\ell,\xi )$  and $\Delta_{\mu\nu}^{F}(\ell)$ will  be
denoted graphically   as  follows (the LV factor $\left(\Delta _{LV}(\vec{\ell}\cdot\vec{\ell})\right)$ is suppressed in the graphs, but its presence is always understood as a factor)~\cite{binosi2009,mavpapav}:

\begin{center}
\begin{picture}(0,20)(100,-15)

\Gluon(-5,-5)(30,-5){2.5}{7} \Photon(160,-5)(195,-5){2}{6}

\Text(35,-5)[l]{\normalsize{$\equiv   i  \Delta_{\mu\nu}(\ell,\xi)$,}}
\Text(200,-5)[l]{\normalsize{$\equiv i \Delta_{\mu\nu}^F(\ell)$.}}

\end{picture}
\end{center}
For the diagrammatic proofs that   will follow, in  addition  to the
propagators   $\Delta_{\mu\nu}(\ell)$  and   $\Delta_{\mu\nu}^F(\ell)$
introduced  above,   we   will need  six   auxiliary  propagator-like
structures, as shown below (again suppression of the LV factors $\left(\Delta _{LV}(\vec{\ell}\cdot \vec{\ell})\right)$ is understood):

\begin{center}
\begin{picture}(0,80)(100,-75)

\Photon(-5,-5)(30,-5){2}{6}
\Photon(-5,-3.5)(30,-3.5){2}{6}
\Text(11,-5)[c]{\rotatebox{19}{\bf{\big /}}}
\Text(10.4,-5)[c]{\rotatebox{19}{\bf{\big /}}}
\Text(14,-5)[c]{\rotatebox{19}{\bf{\big /}}}
\Text(14.6,-5)[c]{\rotatebox{19}{\bf{\big /}}}
\Text(35,-5)[l]{\normalsize{$\equiv\ \frac{\D \ell_\mu
\ell_\nu}{\D \ell^4}$}}

\Photon(-5,-35)(30,-35){2}{6}
\Photon(-5,-33.5)(30,-33.5){2}{6}
\Text(12.2,-35)[c]{\rotatebox{19}{\bf{\big /}}}
\Text(12.8,-35)[c]{\rotatebox{19}{\bf{\big /}}}
\Vertex(-5,-35){2}
\Text(35,-35)[l]{\normalsize{$\equiv\ \frac{\D \ell_\mu}{\D \ell^4}$}}

\Photon(-5,-65)(30,-65){2}{6}
\Photon(-5,-63.5)(30,-63.5){2}{6}
\Vertex(-5,-65){2}
\Vertex(30,-65){2}
\Text(35,-65)[l]{\normalsize{$\equiv\ \frac{\D 1}{\D \ell^4}$}}

\Photon(160,-5)(195,-5){2}{6}
\Text(176,-5)[c]{\rotatebox{19}{\bf{\big /}}}
\Text(175.4,-5)[c]{\rotatebox{19}{\bf{\big /}}}
\Text(179,-5)[c]{\rotatebox{19}{\bf{\big /}}}
\Text(179.6,-5)[c]{\rotatebox{19}{\bf{\big /}}}
\Text(200,-5)[l]{$\equiv\ \frac{\D \ell_\mu
\ell_\nu}{\D \ell^2}$}

\Photon(160,-35)(195,-35){2}{6}
\Vertex(160,-35){2}
\Text(200,-35)[l]{\normalsize{$\equiv\ \frac{\D \ell_\mu}{\D \ell^2}$}}
\Text(177.2,-35)[c]{\rotatebox{19}{\bf{\big /}}}
\Text(177.8,-35)[c]{\rotatebox{19}{\bf{\big /}}}

\Photon(160,-65)(195,-65){2}{6}
\Vertex(160,-65){2}
\Vertex(195,-65){2}
\Text(200,-65)[l]{\normalsize{$\equiv\ \frac{\D 1}{\D \ell^2}$}}

\end{picture}
\end{center}
All of these six structures will arise from  algebraic manipulations of the
original $\Delta_{\mu\nu}(\ell)$. For example, in terms of the above notation
we have the following simple relation (we will set
$\lambda  \equiv \xi-1$):

\begin{center}
\begin{picture}(0,10)(60,20)

\Gluon(5,25)(40,25){2.5}{7}
\Text(45,25)[l]{\normalsize{$\equiv$}}
\Photon(60,25)(95,25){2}{6}
\Text(100,25)[l]{\normalsize{$+\ \lambda$}}
\Photon(120,25.75)(155,25.75){2}{6}
\Photon(120,24.25)(155,24.25){2}{6}
\Text(136,25)[c]{\rotatebox{19}{\bf{\big /}}}
\Text(135.4,25)[c]{\rotatebox{19}{\bf{\big /}}}
\Text(139,25)[c]{\rotatebox{19}{\bf{\big /}}}
\Text(139.6,25)[c]{\rotatebox{19}{\bf{\big /}}}

\end{picture}
\end{center}
We next turn to the study of the gauge-dependence of the fermion self-energy
in our LV QED (but of course the discussion parallels that of electron in QED, quarks in QCD~\cite{binosi2009}). First we oberve that, in view of the fact that in the theory (\ref{bare}) of \cite{alexandre} the fermion part of the action retains its form in Lorentz Invariant QED,
the inverse electron propagator  of order $n$
in the perturbative expansion has the standard form:
\begin{equation}
S_n^{-1}(p,\xi) = \pslush -m - \Sigma^{(n)}(p,\xi)
\end{equation}
where $\Sigma^{(n)} (p,\xi)$ is the $n-th$ order  self-energy. Clearly $
\Sigma^{(0)} = 0$, and $S_0^{-1}(p) = \pslush -m$. The quantity $\Sigma^{(n)}
(p,\xi)$ depends  explicitly on $\xi$ already for $n=1$. In particular

\be
\Sigma^{(1)}(p,\xi) = \int  [d\ell]
\gamma^{\mu} S_0(p+\ell) \gamma^{\nu}
\Delta_{\mu\nu}(\ell,\xi) =
\Sigma^{(1)}_F (p)
+ \lambda \Sigma^{(1)}_L (p)
\label{STOT}
\ee
with $[d\ell] \equiv g^2 \frac{d^4 \ell}{(2\pi)^4}$ and
\begin{equation}
\Sigma^{(1)}_F (p) \equiv \Sigma^{(1)}(p,1) =
\int  [d\ell]
\gamma^{\mu} S_0 (p+\ell) \gamma^{\nu}
\Delta_{\mu\nu}^F (\ell)
\label{SF}
\end{equation}
and
\begin{eqnarray}
\Sigma^{(1)}_L (p)
&=& - S_0^{-1}(p)
\int  \frac{[d\ell]_{LV}}{\ell^4} \, S_0(p+\ell)\gamma^{\nu} \ell_{\nu}
= - \int \frac{[d\ell]_{LV}}{\ell^4} \,   \ell_{\mu}\gamma^{\mu}
S_0(p+\ell)\,\,  S_0^{-1}(p)
\nonumber\\
&=&
  S_0^{-1}(p) \, \int \frac{[d\ell]_{LV}}{\ell^4} S_0(p+\ell) \,\, S_0^{-1}(p)
 - S_0^{-1}(p) \int  \frac{[d\ell]_{LV}}{\ell^4} .
\label{SL}
\end{eqnarray}
Notice that in the above formulas we have absorbed the LV factors of the propagator (\ref{D}) into the integration measure for brevity
\be
[d\ell]_LV \equiv g^2 \Delta_{LV} (\vec{\ell} \cdot \vec{\ell}) \frac{d^4 \ell}{(2\pi)^4}~, \qquad \Delta_{LV} (\vec{\ell}\cdot \vec{\ell}) \equiv \frac{1}{1 + \frac{\vec{\ell}\cdot \vec{\ell}}{M^2}}~.
\label{LVmeasure}
\ee
The r\^ole of these LV terms as Ultraviolet regulators (as $p \rightarrow \infty$) in graphs with internal photon lines is clear. The ordinary Lorentz-Invariant case is recovered in the limit $M \rightarrow \infty$. In the latter limit of course, $[dl]_{M\rightarrow \infty} \rightarrow [d\ell]$ needs regularization (\emph{e.g}. dimensional~\cite{mavpapav}). In our discussion $M $ is fixed and finite and such a regularization is unnecessary for our discussion below. The quantity $g$ is the gauge coupling ($ g\equiv e$ for ordinary
QED). The subscripts ``F'' and ``L'' stand for
``Feynman'' and  ``Longitudinal'', respectively.  Notice that $\Sigma^{(1)}_L$
is proportional to  $S_0^{-1}(p)$ and thus vanishes ``on-shell''.  The most
direct way to arrive at the results of Eq.(\ref{SL}) is to employ the
fundamental WI
\begin{equation}
\elle=S_0^{-1}(p+\ell)-S_0^{-1}(p),
\label{WIzero}
\end{equation}
which is valid in our case in view of the unaltered form of the fermion sector of the theory (\ref{bare}), as compared to the standard QED.

This WI is triggered every time the longitudinal
momenta of $\Delta_{\mu\nu}(\ell,\xi)$ gets contracted with the appropriate
$\gamma$ matrix appearing in the  vertices.
Diagrammatically, this elementary WI gets translated to (again appropriate LV factors $\Delta_{LV}(\vec{\ell}\cdot\vec{\ell})$ are understood)

\begin{center}
\begin{picture}(0,40)(60,20)

\Line(0,25)(50,25)
\Photon(24.3,50)(24.3,25){2}{5}
\Photon(25.7,50)(25.7,25){2}{5}
\Text(25,37.2)[c]{\rotatebox{-71}{\bf{\big /}}}
\Text(25,37.8)[c]{\rotatebox{-71}{\bf{\big /}}}
\Text(55,25)[l]{$\equiv$}

\Line(70,25)(100,25)
\Photon(100.7,25)(100.7,50){2}{5}
\Photon(99.3,25)(99.3,50){2}{5}
\Vertex(100,25){2}

\Text(120,25)[c]{$-$}

\Line(140,25)(170,25)
\Photon(140.7,25)(140.7,50){-2}{5}
\Photon(139.3,25)(139.3,50){-2}{5}
\Vertex(140,25){2}

\end{picture}
\end{center}
Then, the diagrammatic representation of
Eq.(\ref{STOT}), Eq.(\ref{SF}) and Eq.(\ref{SL}) will be given by

\begin{equation}
\begin{picture}(0,110)(130,-45)

\Line(-15,25)(55,25)
\GlueArc(20,25)(20,0,180){2.5}{9}
\Text(60,25)[l]{$\equiv$}

\Line(70,25)(140,25)
\PhotonArc(105,25)(20,0,180){2}{7.5}
\Text(145,25)[l]{$-\ \lambda$}

\Line(170,25)(230,25)
\PhotonArc(190,25)(20,0,180){2}{7.5}
\PhotonArc(190,25)(18.8,0,180){2}{7.5}
\Text(190.9,44.5)[c]{\rotatebox{19}{\bf{\big /}}}
\Text(190.3,44.5)[c]{\rotatebox{19}{\bf{\big /}}}

\Text(60,-30)[l]{$=$}
\Line(70,-30)(140,-30)
\PhotonArc(105,-30)(20,0,180){2}{7.5}
\Text(145,-30)[l]{$+\ \lambda$}

\Line(170,-30)(230,-30)
\PhotonArc(200,-40)(30,20,160){2}{7.5}
\PhotonArc(200,-40)(28.8,20,160){2}{7.5}
\Vertex(170,-30){2}
\Vertex(230,-30){2}
\Text(237,-30)[l]{$-\ \lambda$}

\Line(270,-30)(320,-30)
\PhotonArc(270,-20)(10,180,540){2}{9}
\PhotonArc(270,-20)(8.8,180,540){2}{9}

\Vertex(270,-30){2}
\Vertex(170,25){2}

\label{GR}
\end{picture}
\end{equation}

This diagrammatic analysis is all one needs to prove the validity of the PT
to our case. Indeed, when considering physical amplitudes,  the characteristic structure of the
longitudinal parts established above   allows for their cancellation against
identical contributions  originating from diagrams which are kinematically
different from fermion self-energies,  such as vertex-graphs or boxes, {\it
without} the need for integration over the internal virtual momenta.  This last
property is important because  in this way the original  kinematical identity
is guaranteed to be maintained; instead, loop integrations generally mix the
various kinematics~\footnote{It is now clear to the reader that it is also this property that guarantees the straightforward application of the PT of Lorentz-Invariant gauge theories to our LV QED (\ref{bare}), as a result of the factorization of the LV terms in the photon propagator.}. Diagrammatically, the action of the WI  is very
distinct:  it always gives rise to \emph{unphysical} effective vertices, {\it i.e.}
vertices
which do not appear in the original Lagrangian; all such vertices   \emph{cancel} in
the full, gauge-invariant amplitude.

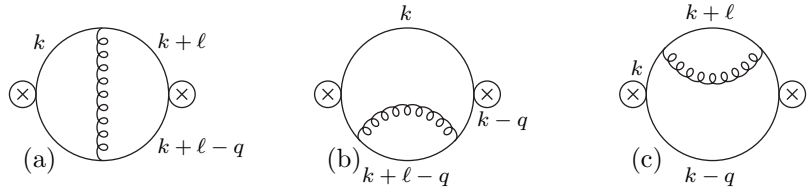
\begin{figure}[t]

\begin{picture}(100,100)(-60,0)

\Text(20,25)[l]{\footnotesize{(a)}}
\Text(135,25)[l]{\footnotesize{(b)}}
\Text(250,25)[l]{\footnotesize{(c)}}

\Text(24,70)[l]{\scriptsize{$k$}}
\Text(71,70)[l]{\scriptsize{$k+\ell$}}
\Text(71,30)[l]{\scriptsize{$k+\ell-q$}}

\Text(165,81)[c]{\scriptsize{$k$}}
\Text(192,40)[l]{\scriptsize{$k-q$}}
\Text(165,19)[c]{\scriptsize{$k+\ell-q$}}

\Text(280,81)[c]{\scriptsize{$k+\ell$}}
\Text(250,60)[l]{\scriptsize{$k$}}
\Text(280,19)[c]{\scriptsize{$k-q$}}

\CArc(20,50)(5,0,360)
\Line(18,52)(22,48)
\Line(22,52)(18,48)
\CArc(50,50)(25,0,360)
\CArc(80,50)(5,0,360)
\Line(78,52)(82,48)
\Line(82,52)(78,48)

\CArc(135,50)(5,0,360)
\Line(133,52)(137,48)
\Line(137,52)(133,48)
\CArc(165,50)(25,0,360)
\CArc(195,50)(5,0,360)
\Line(193,52)(197,48)
\Line(197,52)(193,48)

\CArc(250,50)(5,0,360)
\Line(248,52)(252,48)
\Line(252,52)(248,48)
\CArc(280,50)(25,0,360)
\CArc(310,50)(5,0,360)
\Line(308,52)(312,48)
\Line(312,52)(308,48)

\Gluon(50,25)(50,75){2}{9}

\GlueArc(165,25)(19.2,22.5,157.5){2}{9}

\GlueArc(280,75)(19.2,202.5,337.5){2}{9}

\end{picture}

\caption{One loop diagram contributing to the Lorentz-Violating QED fermion self-energy. The diagram is formally the same as in the Lorentz-invariant case due to the specific form of the action (\ref{bare}).}

\label{fig1}
\end{figure}

To actually pursue these special
cancellations explicitly one may choose among a variety
of gauge invariant quantities.  For our interest, such an example is provided by the fermion-current
correlation function  $I_{\mu\nu}$ defined as (in momentum space)
\be
I_{\mu\nu}(q) = i \int d^{4}x e^{iq\cdot x}
\langle 0  | T \left[J_{\mu}(x) J_{\nu}(0) \right] |0 \rangle
 = (g_{\mu\nu} q^2 - q_{\mu}q_{\nu}) I(q^2)I_{LV} (\vec{q}\cdot \vec{q})\, ,
\label{IQED}
\ee
where the current $J_{\mu} (x)$ is given by $J_{\mu}(x) =\ :\!\bar{\psi} (x)
\gamma_{\mu} \psi (x)\!:$. $I_{\mu\nu}(q)$  coincides with the photon
vacuum polarization of QED. The quantity $q$ denotes four momentum, $\vec{q}$ the spatial part, and
the factoring out of the LV term $I(\vec{q}\cdot \vec{q}) $ is a consequence of the photon propagator
(\ref{bare}), which maintains gauge invariance (and hence masslessness of the photon) in the presence of LV~\cite{alexandre}.

To see explicitly the mechanism enforcing  these cancellations, we first  consider the one-loop photonic  corrections to the quantity  $I_{\mu\nu}$. Clearly such corrections
is GFP-independent, since the current $J_{\mu}(x)$ is invariant under both the
$U(1)$ gauge transformations. The relevant diagrams are those shown in Fig.\ref{fig1}.

To see the
appearance of the unphysical vertices,
we carry out the manipulations presented
in Eq.(\ref{STOT}), Eq.(\ref{SF}), and Eq.(\ref{SL}), or, equivalently, in
 Eq.(\ref{GR}),
this time  embedded inside $I_{\mu\nu}(q)$. Thus, from  diagrams (b) and
(c) we arrive at
\begin{center}
\begin{picture}(100,50)(30,25)

\Text(-55,50)[l]{(b)+(c) $\to\ 2\lambda$}

\CArc(20,50)(5,0,360)
\Line(18,52)(22,48)
\Line(22,52)(18,48)
\CArc(50,50)(25,0,360)
\CArc(80,50)(5,0,360)
\Line(78,52)(82,48)
\Line(82,52)(78,48)
\PhotonArc(50,25)(19.2,22.5,157.5){2}{9}
\PhotonArc(50,25)(18,22.5,157.5){2}{9}
\Text(48.5,44.2)[c]{\rotatebox{19}{\bf{\big /}}}
\Text(47.9,44.2)[c]{\rotatebox{19}{\bf{\big /}}}
\Text(51.5,44.2)[c]{\rotatebox{19}{\bf{\big /}}}
\Text(52.1,44.2)[c]{\rotatebox{19}{\bf{\big /}}}

\Text(90,50)[l]{$=-\,2\lambda$}
\hspace{-1.3 cm}

\CArc(160,50)(5,0,360)
\Line(158,52)(162,48)
\Line(162,52)(158,48)
\CArc(190,50)(25,0,360)
\CArc(220,50)(5,0,360)
\Line(218,52)(222,48)
\Line(222,52)(218,48)
\PhotonArc(215,25)(25.7,90,180){-2}{7.5}
\PhotonArc(215,25)(24.3,90,180){-2}{7.5}
\Vertex(215,50){2}
\Text(196,41)[c]{\rotatebox{65}{\bf{\big /}}}
\Text(196.45,41.45)[c]{\rotatebox{65}{\bf{\big /}}}

\end{picture}
\end{center}
Again, the reader should understand the presence of LV factors $\Delta_{LV}(\vec{k}\vec{k})$
in the appropriate parts of the graphs, but they do not affect the general arguments on the GFP dependence cancellation.

We thus see that since the action of the
elementary WI of  Eq.(\ref{WIzero}) amounts to the cancellation of internal
propagators,
its diagrammatic
consequence
is that of introducing an unphysical effective vertex,
describing an interaction of the form $\gamma\gamma \bar{\psi} \psi$,
where $\gamma$ denotes a photon field.  This type of vertex
may be depicted by means of a Feynman rule of the form

\begin{center}
\begin{picture}(0,35)(20,10)

\CArc(20,10)(5,0,360)
\Line(18,12)(22,8)
\Line(22,12)(18,8)

\Line(-5,15)(45,15)
\Gluon(20,15)(20,40){2}{5}
\Vertex(20,15){2}

\Text(50,15)[l]{$\equiv i\gamma_\mu$}
\Text(30,10)[c]{\footnotesize{$\mu$}}

\end{picture}
\end{center}
being $\mu$ the index of the external current.

To see how the above unphysical contributions cancel inside  $I_{\mu\nu}$
we turn to diagram (a). The action of
the WI may be
translated to the following diagrammatic picture

\begin{center}
\begin{picture}(100,50)(85,25)

\Text(-30,50)[l]{(a) $\to\ \lambda$}

\CArc(20,50)(5,0,360)
\Line(18,52)(22,48)
\Line(22,52)(18,48)
\CArc(50,50)(25,0,360)
\CArc(80,50)(5,0,360)
\Line(78,52)(82,48)
\Line(82,52)(78,48)
\Photon(49.25,25)(49.25,75){2}{9}
\Photon(50.75,25)(50.75,75){2}{9}
\Text(50,51.5)[c]{\rotatebox{-71}{\bf{\big /}}}
\Text(50,52.1)[c]{\rotatebox{-71}{\bf{\big /}}}
\Text(50,48.5)[c]{\rotatebox{-71}{\bf{\big /}}}
\Text(50,47.9)[c]{\rotatebox{-71}{\bf{\big /}}}
\Text(90,50)[l]{$=\ \ \lambda$}
\hspace{-1.5 cm}

\Text(155,25)[lb]{\footnotesize{$(\alpha)$}}

\CArc(160,50)(5,0,360)
\Line(158,52)(162,48)
\Line(162,52)(158,48)
\CArc(190,50)(25,0,360)
\CArc(220,50)(5,0,360)
\Line(218,52)(222,48)
\Line(222,52)(218,48)
\PhotonArc(215,25)(25.7,90,180){-2}{7.5}
\PhotonArc(215,25)(24.3,90,180){-2}{7.5}
\Vertex(215,50){2}
\Text(196,41)[c]{\rotatebox{65}{\bf{\big /}}}
\Text(196.45,41.45)[c]{\rotatebox{65}{\bf{\big /}}}
\Text(230,50)[l]{$-\ \ \lambda$}
\hspace{-1.6 cm}

\Text(295,25)[lb]{\footnotesize{$(\beta)$}}

\CArc(300,50)(5,0,360)
\Line(298,52)(302,48)
\Line(302,52)(298,48)
\CArc(330,50)(25,0,360)
\CArc(360,50)(5,0,360)
\Line(358,52)(362,48)
\Line(362,52)(358,48)
\PhotonArc(305,25)(25.7,0,90){2}{7.5}
\PhotonArc(305,25)(24.3,0,90){2}{7.5}
\Vertex(305,50){2}
\Text(324,41)[c]{\rotatebox{-25}{\bf{\big /}}}
\Text(323.6,41.4)[c]{\rotatebox{-25}{\bf{\big /}}}

\end{picture}
\end{center}
It is then elementary to establish that the two diagrams on the
right-hand side
of the
above diagrammatic  equation add up.

Summing up the two equations above, it is clear how  the gauge dependent part
of the one loop amplitude cancel completely.
Having proved that the GFP-dependent contributions coming from the
original graphs containing $\Sigma^{(1)}(p,\xi)$, {\it i.e.}
Fig.\ref{fig1}(b) and Fig.\ref{fig1}(c),  cancel exactly against
equal but opposite {\it propagator-like} contributions coming from
Fig.\ref{fig1}(a), one is left with the ``pure'' GFP-independent
one-loop fermion self-energy, $\widehat{\Sigma}^{(1)}(p)$. Clearly,
it coincides with the $\Sigma^{(1)}_F (p)$ of Eq.(\ref{SF}), i.e.
\begin{equation}
\widehat{\Sigma}^{(1)}(p) = \Sigma^{(1)}_F (p).
\label{GFPI}
\end{equation}

This implies that in any physically measurable quantities, such as scattering amplitudes or dynamically generated masses, of interest at hand, the only relevant part of the self-energy that will contribute
is the one (\ref{GFPI}) associated with the Feynman gauge $\xi = 1$. With this in mind, the relevant computations in the SD analysis for dynamical mass generation, for instance, are simplified by going to this gauge and dropping the longitudinal parts of the photon propagator~\cite{binosi2009,mavpapav}, for reasons explained above.


\begin{thebibliography}{99}

\bibitem{alexandre} J.~Alexandre,
  %``Dynamical mass generation in Lorentz-violating QED,''
  arXiv:1009.5834 [hep-ph].
  %%CITATION = ARXIV:1009.5834;%%



\bibitem{kostel} D.~Colladay and V.~A.~Kostelecky,
  %``CPT violation and the standard model,''
  Phys.\ Rev.\  D {\bf 55}, 6760 (1997)
  [arXiv:hep-ph/9703464];
  %%CITATION = PHRVA,D55,6760;%%
V.~A.~Kostelecky and S.~Samuel,
  %``Gravitational Phenomenology In Higher Dimensional Theories And Strings,''
  Phys.\ Rev.\  D {\bf 40}, 1886 (1989);
  %%CITATION = PHRVA,D40,1886;%%
 V.~A.~Kostelecky,
  %``Perspectives on Lorentz and CPT Violation,''
  arXiv:0802.0581 [gr-qc], and references therein.
  %%CITATION = ARXIV:0802.0581;%%



\bibitem{horava} P.~Horava,
  %``Quantum Gravity at a Lifshitz Point,''
  Phys.\ Rev.\  D {\bf 79}, 084008 (2009)
  [arXiv:0901.3775 [hep-th]].
  %%CITATION = PHRVA,D79,084008;%%

\bibitem{sotiriou} see for instance, T.~P.~Sotiriou, M.~Visser and S.~Weinfurtner,
  %``Phenomenologically viable Lorentz-violating quantum gravity,''
  Phys.\ Rev.\ Lett.\  {\bf 102}, 251601 (2009)
  [arXiv:0904.4464 [hep-th]].
  %%CITATION = PRLTA,102,251601;%%

\bibitem{posp} M.~Pospelov and Y.~Shang,
  %``On Lorentz violation in Horava-Lifshitz type theories,''
  arXiv:1010.5249 [hep-th].
  %%CITATION = ARXIV:1010.5249;%%

\bibitem{visser} M.~Visser,
  %``Lorentz symmetry breaking as a quantum field theory regulator,''
  Phys.\ Rev.\  D {\bf 80}, 025011 (2009)
  [arXiv:0902.0590 [hep-th]] and references therein.
  %%CITATION = PHRVA,D80,025011;%%




\bibitem{dynmas} See, for instance: D.~Anselmi,
  %``Standard Model Without Elementary Scalars And High Energy Lorentz
  %Violation,''
  Eur.\ Phys.\ J.\  C {\bf 65}, 523 (2010)
  [arXiv:0904.1849 [hep-ph]];
  %%CITATION = EPHJA,C65,523;%%
D.~Anselmi and E.~Ciuffoli,
  %``Renormalization Of High-Energy Lorentz Violating Four Fermion Models,''
  Phys.\ Rev.\  D {\bf 81}, 085043 (2010)
  [arXiv:1002.2704 [hep-ph]];
  %%CITATION = PHRVA,D81,085043;%%
J.~Alexandre, K.~Farakos, P.~Pasipoularides and A.~Tsapalis,
  %``Schwinger-Dyson Approach For A Lifshitz-Type Yukawa Model,''
  Phys.\ Rev.\  D {\bf 81}, 045002 (2010)
  [arXiv:0909.3719 [hep-th]];
  %%CITATION = PHRVA,D81,045002;%%
J.~Alexandre, N.~E.~Mavromatos and D.~Yawitch,
  %``Emergent Lorentz symmetry and Dynamical Mass Generation in the Infra-Red
  %limit of a Lifshitz-type Yukawa model,''
  arXiv:1009.4811 [hep-ph].
  %%CITATION = ARXIV:1009.4811;%%






\bibitem{dfoam} J.~R.~Ellis, N.~E.~Mavromatos and D.~V.~Nanopoulos,
%``Quantum-gravitational diffusion and stochastic fluctuations in the
 %velocity of light,''
 Gen.\ Rel.\ Grav.\ \textbf{32}, 127 (2000); %%CITATION = GRGVA,32,127;%%
%``A microscopic recoil model for light-cone fluctuations in quantum
 %gravity,''
 Phys.\ Rev.\ D \textbf{61}, 027503 (2000); %%CITATION = PHRVA,D61,027503;%%
 %``Dynamical formation of horizons in recoiling D-branes,''
 Phys.\ Rev.\ D \textbf{62}, 084019 (2000).

\bibitem{westmuckett}  J.~R.~Ellis, N.~E.~Mavromatos
and M.~Westmuckett, %``A supersymmetric D-brane model of space-time foam,''
Phys.\ Rev.\ D \textbf{70}, 044036 (2004); %%CITATION = PHRVA,D70,044036;%%
\emph{ibid.} \textbf{71}, 106006 (2005)~.

\bibitem{emnnewuncert} J.~R.~Ellis, N.~E.~Mavromatos and D.~V.~Nanopoulos,
%``Derivation of a Vacuum Refractive Index in a Stringy Space-Time Foam
 %Model,''
 Phys.\ Lett.\ B \textbf{665}, 412 (2008); %%CITATION = PHLTA,B665,412;%%
%``D-Foam Phenomenology: Dark Energy, the Velocity of Light and a Possible
 %D-Void,''
 arXiv:0912.3428 {[}astro-ph.CO{]}; %%CITATION = ARXIV:0912.3428;%%
J.~Ellis, N.~E.~Mavromatos and D.~V.~Nanopoulos,
  %``Comments on Ultra-High-Energy Photons and D-Foam models,''
  Phys.\ Lett.\  B {\bf 694}, 61 (2010)
  [arXiv:1004.4167 [astro-ph.HE]].
  %%CITATION = PHLTA,B694,61;%%


\bibitem{li} T.~Li, N.~E.~Mavromatos, D.~V.~Nanopoulos and D.~Xie,
%``Time Delays of Strings in D-particle Backgrounds and Vacuum Refractive
 %Indices,''
 Phys.\ Lett.\ B \textbf{679}, 407 (2009). %%CITATION = PHLTA,B679,407;%%



\bibitem{review} For a recent review on this topic see: N.~E.~Mavromatos,
  %``String Quantum Gravity, Lorentz-Invariance Violation and Gamma-Ray
  %Astronomy,''
  arXiv:1010.5354 [hep-th]: invited review to appear in Int. J. Mod. Phys. A,  and references therein.
  %%CITATION = ARXIV:1010.5354;%%


\bibitem{vergou} N.~E.~Mavromatos, S.~Sarkar and A.~Vergou,
  %``Stringy Space-Time Foam, Finsler-like Metrics and Dark Matter Relics,''
  arXiv:1009.2880 [hep-th].
  %%CITATION = ARXIV:1009.2880;%%


\bibitem{szabo}  N.~E.~Mavromatos and R.~J.~Szabo, %``Matrix D-brane dynamics, logarithmic operators and quantization of
 %noncommutative spacetime,''
 Phys.\ Rev.\ D \textbf{59}, 104018 (1999) {[}arXiv:hep-th/9808124{]};
%%CITATION = PHRVA,D59,104018;%%
see also: J.~R.~Ellis, N.~E.~Mavromatos and D.~V.~Nanopoulos,
  %``A String scenario for inflationary cosmology,''
  Mod.\ Phys.\ Lett.\  A {\bf 10}, 1685 (1995)
  [arXiv:hep-th/9503162];
  %%CITATION = MPLAE,A10,1685;%%
G.~Amelino-Camelia, J.~R.~Ellis, N.~E.~Mavromatos and D.~V.~Nanopoulos,
  %``On the space-time uncertainty relations of Liouville strings and D
  %branes,''
  Mod.\ Phys.\ Lett.\  A {\bf 12}, 2029 (1997)
  [arXiv:hep-th/9701144].
  %%CITATION = MPLAE,A12,2029;%%

%\cite{Cornwall:1982zr}
\bibitem{Cornwall:1982zr}
J.~M.~Cornwall,
%``Dynamical Mass Generation In Continuum QCD,''
Phys.\ Rev.\ D {\bf 26}, 1453 (1982);
%%CITATION = PHRVA,D26,1453;%%
J.~M.~Cornwall and J.~Papavassiliou,
%``Gauge Invariant three-gluon Vertex In QCD,''
Phys.\ Rev.\ D {\bf 40}, 3474 (1989).
%%CITATION = PHRVA,D40,3474;%%




\bibitem{binosi2009} D.~Binosi and J.~Papavassiliou,
  %``Pinch Technique: Theory and Applications,''
  Phys.\ Rept.\  {\bf 479}, 1 (2009)
  [arXiv:0909.2536 [hep-ph]] and references therein.
  %%CITATION = PRPLC,479,1;%%



\bibitem{B}
V.~P.~Gusynin, V.~A.~Miransky and I.~A.~Shovkovy,
  %``Dynamical chiral symmetry breaking by a magnetic field in QED,''
  Phys.\ Rev.\  D {\bf 52} (1995) 4747
  [arXiv:hep-ph/9501304].
  %%CITATION = PHRVA,D52,4747;%%

\bibitem{gms} V.~P.~Gusynin, V.~A.~Miransky and I.~A.~Shovkovy,
  %``Physical Gauge In The Problem Of Dynamical Chiral Symmetry Breaking In QED
  %In A Magnetic Field,''
  Found.\ Phys.\  {\bf 30}, 349 (2000) and references therein.
  %%CITATION = FNDPA,30,349;%%







\bibitem{miransky}
V.~A.~Miransky,
``Dynamical symmetry breaking in quantum field theories,''
%\href{http://www.slac.stanford.edu/spires/find/hep/www?irn=3034437}{SPIRES entry}
{\it  Singapore, Singapore: World Scientific (1993) 533 p}


\bibitem{mavpapav}  N.~E.~Mavromatos, J.~Papavassiliou,
  %``Novel phases and old puzzles in QED3 and related models,''
Recent. Res. Devel. Phys. 5, 369-415 (TRN, India 2004)
[arXiv:cond-mat/0311421], invited article.
%%CITATION = COND-MAT 0311421;%%

\bibitem{RS} L.~Randall and R.~Sundrum,
  %``A large mass hierarchy from a small extra dimension,''
  Phys.\ Rev.\ Lett.\  {\bf 83}, 3370 (1999)
  [arXiv:hep-ph/9905221].
  %%CITATION = PRLTA,83,3370;%%

\bibitem{MR} N.~E.~Mavromatos and J.~Rizos,
  %``Exact solutions and the cosmological constant problem in dilatonic  domain
  %wall higher-curvature string gravity,''
  Int.\ J.\ Mod.\ Phys.\  A {\bf 18}, 57 (2003)
  [arXiv:hep-th/0205299];
  %%CITATION = IMPAE,A18,57;%%
%``String-Inspired Higher-Curvature Terms and the Randall-Sundrum Scenario,''
  Phys.\ Rev.\  D {\bf 62}, 124004 (2000)
  [arXiv:hep-th/0008074].
  %%CITATION = PHRVA,D62,124004;%%


\bibitem{lcft} For a very partial but somewhat relevant works to our discussion here see: V.~Gurarie,
  %``Logarithmic operators in conformal field theory,''
  Nucl.\ Phys.\  B {\bf 410}, 535 (1993)
  [arXiv:hep-th/9303160];
  %%CITATION = NUPHA,B410,535;%%
J.~S.~Caux, I.~I.~Kogan and A.~M.~Tsvelik,
  %``Logarithmic operators and hidden continuous symmetry in critical disordered
  %models,''
  Nucl.\ Phys.\  B {\bf 466}, 444 (1996)
  [arXiv:hep-th/9511134];
  %%CITATION = NUPHA,B466,444;%%
I.~I.~Kogan and N.~E.~Mavromatos,
  %``World-Sheet Logarithmic Operators and Target Space Symmetries in String
  %Theory,''
  Phys.\ Lett.\  B {\bf 375}, 111 (1996)
  [arXiv:hep-th/9512210].
  %%CITATION = PHLTA,B375,111;%%
 M.~A.~I.~Flohr,
  %``Logarithmic conformal field theory - or - how to compute a torus  amplitude
  %on the sphere,''
  arXiv:hep-th/0407003.
  %%CITATION = HEP-TH/0407003;%%
J.~Fuchs, S.~Hwang, A.~M.~Semikhatov and I.~Y.~Tipunin,
  %``Nonsemisimple fusion algebras and the Verlinde formula,''
  Commun.\ Math.\ Phys.\  {\bf 247}, 713 (2004)
  [arXiv:hep-th/0306274].
  %%CITATION = CMPHA,247,713;%%
M.~S.~Movahed, M.~Saadat and M.~Reza Rahimi Tabar,
  %``The O(n) model in the n --> 0 limit (self-avoiding-walks) and  logarithmic
  %conformal field theory,''
  Nucl.\ Phys.\  B {\bf 707}, 405 (2005)
  [arXiv:cond-mat/0409486], and references therein.
  %%CITATION = NUPHA,B707,405;%%


\bibitem{recoil} I.~I.~Kogan, N.~E.~Mavromatos and J.~F.~Wheater,
  %``D-brane recoil and logarithmic operators,''
  Phys.\ Lett.\  B {\bf 387}, 483 (1996);
  %%CITATION = PHLTA,B387,483;%%
J.~R.~Ellis, N.~E.~Mavromatos and D.~V.~Nanopoulos,
  %``D-brane recoil mislays information,''
  Int.\ J.\ Mod.\ Phys.\  A {\bf 13}, 1059 (1998).
  %%CITATION = IMPAE,A13,1059;%%

\bibitem{finsler} See, for instance: D. Bao, S.~S.~Chern and Z.~Shen, \emph{An introduction
to Finsler Geometry} (Springer-Verlag (NY, 2000)).
In the context of D-particle foam, such Finsler-type metrics have been first derived in:
J.~R.~Ellis, N.~E.~Mavromatos and D.~V.~Nanopoulos,
  %``D-brane recoil mislays information,''
  Int.\ J.\ Mod.\ Phys.\  A {\bf 13}, 1059 (1998)
  [arXiv:hep-th/9609238];
  %%CITATION = IMPAE,A13,1059;%%
For a short recent review on this topic see: N.~E.~Mavromatos,
  %``Lorentz Invariance Violation from String Theory,''
  PoS {\bf QG-PH}, 027 (2007)
  [arXiv:0708.2250 [hep-th]] and references therein.
  %%CITATION = POSCI,QG-PH,027;%%
Also Finsler metrics in string theory, but in a different context, have been previously suggested in : S.~I.~Vacaru,
  %``(Non)commutative Finsler geometry from string / M-theory,''
  arXiv:hep-th/0211068;
  %%CITATION = HEP-TH/0211068;%%
S.~I.~Vacaru,
  %``Generalized Finsler geometry in Einstein, string and metric-affine
  %gravity,''
  arXiv:hep-th/0310132;
  %%CITATION = HEP-TH/0310132;%%
In a field theory context, such metrics have been discussed, among other works, in:
 G.~Y.~Bogoslovsky,
  %``Some physical displays of the space anisotropy relevant to the feasibility
  %of its being detected at a laboratory,''
  arXiv:0706.2621 [gr-qc];
  %%CITATION = ARXIV:0706.2621;%%
  %``Rapidities and observable 3-velocities in the flat Finslerian event space
  %with entirely broken 3D isotropy,''
  arXiv:0712.1718 [hep-th].
  %%CITATION = ARXIV:0712.1718;%%
G.~W.~Gibbons, J.~Gomis and C.~N.~Pope,
  %``General Very Special Relativity is Finsler Geometry,''
  Phys.\ Rev.\  D {\bf 76}, 081701 (2007)
  [arXiv:0707.2174 [hep-th]];
  %%CITATION = PHRVA,D76,081701;%%
  A.~P.~Kouretsis, M.~Stathakopoulos and P.~C.~Stavrinos,
  %``The General Very Special Relativity in Finsler Cosmology,''
  arXiv:0810.3267 [gr-qc];
  %%CITATION = ARXIV:0810.3267;%%
L.~Sindoni,
  %``The Higgs mechanism in Finsler spacetimes,''
  Phys.\ Rev.\  D {\bf 77}, 124009 (2008)
  [arXiv:0712.3518 [gr-qc]];
  %%CITATION = PHRVA,D77,124009;%%
M.~Anastasiei and S.~I.~Vacaru,
  %``Fedosov Quantization of Lagrange-Finsler and Hamilton-Cartan Spaces and
  %Einstein Gravity Lifts on (Co) Tangent Bundles,''
  J.\ Math.\ Phys.\  {\bf 50}, 013510 (2009)
  [arXiv:0710.3079 [math-ph]].
  %%CITATION = JMAPA,50,013510;%%
In the context of generic phenomenological models of non-standard dispersion relations in quantum gravity and LV Horava-Lifshitz theories, see:
F.~Girelli, S.~Liberati and L.~Sindoni,
  %``Phenomenology of quantum gravity and Finsler geometry,''
  Phys.\ Rev.\  D {\bf 75}, 064015 (2007)
  [arXiv:gr-qc/0611024];
  %%CITATION = PHRVA,D75,064015;%%
J.~Magueijo and L.~Smolin,
  %``Gravity's Rainbow,''
  Class.\ Quant.\ Grav.\  {\bf 21}, 1725 (2004)
  [arXiv:gr-qc/0305055].
  %%CITATION = CQGRD,21,1725;%%
J.~Skakala and M.~Visser,
  %``Birefringence in pseudo-Finsler spacetimes,''
  arXiv:0810.4376 [gr-qc];
  %%CITATION = ARXIV:0810.4376;%%
S.~I.~Vacaru,
  %``Modified Dispersion Relations in Horava-Lifshitz Gravity and Finsler Brane
  %Models,''
  arXiv:1010.5457 [math-ph].
  %%CITATION = ARXIV:1010.5457;%%


\bibitem{seibergwitten} N.~Seiberg and E.~Witten,
  %``String theory and noncommutative geometry,''
  JHEP {\bf 9909}, 032 (1999)
  [arXiv:hep-th/9908142].
  %%CITATION = JHEPA,9909,032;%%



\bibitem{sussk1} N.~Seiberg, L.~Susskind and N.~Toumbas,
  %``Space/time non-commutativity and causality,''
  JHEP {\bf 0006}, 044 (2000).
  %%CITATION = JHEPA,0006,044;%%


\bibitem{sarkar}  N.E.~Mavromatos and Sarben Sarkar,
%``Liouville decoherence in a model of flavour oscillations in the  presence
%  of dark energy,''
 Physical\ Review\ D \textbf{72}, 065016 (2005) {[}arXiv:hep-th/0506242{]}.
%%CITATION = HEP-TH 0506242;%%




\bibitem{tseytlin}  see, for instance: A.~A.~Tseytlin,
  %``Born-Infeld action, supersymmetry and string theory,''
  arXiv:hep-th/9908105.
  %%CITATION = HEP-TH/9908105;%%


\bibitem{szabosusy} N.~E.~Mavromatos, R.~J.~Szabo,
  %``The Neveu-Schwarz and Ramond algebras of logarithmic superconformal field theory,''
  JHEP {\bf 0301}, 041 (2003).
  [hep-th/0207273];
%``D-brane dynamics and logarithmic superconformal algebras,''
  JHEP {\bf 0110}, 027 (2001).
  [hep-th/0106259].

\bibitem{lpole} See, for example, for a lattice attempt to address this issue: M.~Gockeler, R.~Horsley, V.~Linke {\it et al.},
  %``Is there a Landau pole problem in QED?,''
  Phys.\ Rev.\ Lett.\  {\bf 80}, 4119-4122 (1998).
  [hep-th/9712244].


\bibitem{giundi} See, for instance: C.~Giunti, A.~Studenikin,
  %``Neutrino electromagnetic properties,''
  Phys.\ Atom.\ Nucl.\  {\bf 72}, 2089-2125 (2009).
  [arXiv:0812.3646 [hep-ph]], and references therein.


\bibitem{sarkarbeny} J.~Bernabeu, N.~E.~Mavromatos, J.~Papavassiliou,
  %``Novel type of CPT violation for correlated EPR states,''
  Phys.\ Rev.\ Lett.\  {\bf 92}, 131601 (2004).
  [hep-ph/0310180];
J.~Bernabeu, N.~E.~Mavromatos, S.~Sarkar,
  %``Decoherence induced CPT violation and entangled neutral mesons,''
  Phys.\ Rev.\  {\bf D74}, 045014 (2006).
  [hep-th/0606137];
N.~E.~Mavromatos,
  %``Decoherence and CPT Violation in a Stringy Model of Space-Time Foam,''
  Found.\ Phys.\  {\bf 40}, 917-960 (2010).
  [arXiv:0906.2712 [hep-th]] and references therein.


\bibitem{bar} G.~Barenboim, N.~E.~Mavromatos,
  %``CPT violating decoherence and LSND: A Possible window to Planck scale physics,''
  JHEP {\bf 0501}, 034 (2005).
  [hep-ph/0404014];
%``Decoherent neutrino mixing, dark energy and matter-antimatter asymmetry,''
  Phys.\ Rev.\  {\bf D70}, 093015 (2004).
  [hep-ph/0406035];
G.~Barenboim, N.~E.~Mavromatos, S.~Sarkar {\it et al.},
  %``Quantum decoherence and neutrino data,''
  Nucl.\ Phys.\  {\bf B758}, 90-111 (2006).
  [hep-ph/0603028].


%\cite{Cornwall:1974vz}
\bibitem{Cornwall:1974vz}
J.~M.~Cornwall, R.~Jackiw and E.~Tomboulis,
%``Effective Action For Composite Operators,''
Phys.\ Rev.\ D {\bf 10}, 2428 (1974);
%%CITATION = PHRVA,D10,2428;%%
W.~J.~Marciano and H.~Pagels,
%``Quantum Chromodynamics: A Review,''
Phys.\ Rept.\  {\bf 36}, 137 (1978).
%%CITATION = PRPLC,36,137;%%







%\cite{Mavromatos:1999jf}
\bibitem{Mavromatos:1999jf}
N.~E.~Mavromatos and J.~Papavassiliou,
%``Non-linear dynamics in QED(3) and non-trivial infrared structure,''
Phys.\ Rev.\ D {\bf 60}, 125008 (1999).
%[arXiv:hep-th/9904046].
%%CITATION = HEP-TH 9904046;%%}.



%\cite{Sauli:2002tk}
\bibitem{Sauli:2002tk}
V.~Sauli,
%``Minkowski solution of Dyson-Schwinger equations in momentum subtraction scheme,''
JHEP {\bf 0302}, 001 (2003)
[arXiv:hep-ph/0209046].
%%CITATION = HEP-PH 0209046;%%



%\cite{Binosi:2002ft}
\bibitem{Binosi:2002ft}
D.~Binosi and J.~Papavassiliou,
%``The pinch technique to all orders,''
Phys.\ Rev.\ D {\bf 66}, 111901 (2002)
[arXiv:hep-ph/0208189];
%%CITATION = HEP-PH 0208189;%%
%``Pinch technique self-energies and vertices to all orders in perturbation
  %theory,''
  J.\ Phys.\ G {\bf 30}, 203 (2004)
  [arXiv:hep-ph/0301096].
  %%CITATION = JPHGB,G30,203;%%





\end{thebibliography}
\end{document}